\pdfoutput=1
\documentclass[11pt,twoside,a4paper,cmspaper,final,collab]{cms-tdr}
\def\svnVersion{e17ff13}\def\svnDate{2024/07/03}\def\cmsCernNoTag{CERN-EP-2023-220}\def\cmsCernDate{\today}\def\cmsMessage{Published in Physical Review Letters as \href{http://dx.doi.org/10.1103/PhysRevLett.133.011801}{\doi{10.1103/PhysRevLett.133.011801}.}}
\begin{document}\cmsNoteHeader{EXO-22-008}

\newcommand{\mjjj}{\ensuremath{m_{\mathrm{jjj}}}\xspace}
\newcommand{\rhom}{\ensuremath{\rho_{\mathrm{m}}}\xspace}
\newcommand{\fb} {\mbox{\ensuremath{\,\text{fb}}}\xspace}
\newcommand{\pb} {\mbox{\ensuremath{\,\text{pb}}}\xspace}
\DeclareRobustCommand{\PX}{{\HepParticle{X}{}{}}\Xspace}
\DeclareRobustCommand{\PY}{{\HepParticle{Y}{}{}}\Xspace}
\DeclareRobustCommand{\PGKK}{{\HepParticle{g}{KK}{}}\Xspace}

\ifthenelse{\boolean{cms@external}}{\providecommand{\cmsLeft}{upper\xspace}}{\providecommand{\cmsLeft}{left\xspace}}
\ifthenelse{\boolean{cms@external}}{\providecommand{\cmsRight}{lower\xspace}}{\providecommand{\cmsRight}{right\xspace}}

\ifthenelse{\boolean{cms@external}}{\providecommand{\suppMaterial}{the supplemental material~\cite{supp}}}{\providecommand{\suppMaterial}{Appendix~\ref{app:suppMat}\nocite{supp}\xspace}}
\cmsNoteHeader{EXO-22-008}
\title{Search for narrow trijet resonances in proton-proton collisions at \texorpdfstring{$\sqrt{s} = 13\TeV$}{sqrt(s) = 13 TeV}}
\date{\today}

\abstract{
     The first search for singly produced narrow resonances decaying to three well-separated hadronic jets is presented. The search uses proton-proton collision data corresponding to an integrated luminosity of 138\fbinv at $\sqrt{s}= 13\TeV$, collected at the CERN LHC. No significant deviations from the background predictions are observed between 1.75 and 9.00\TeV. The results provide the first mass limits on a right-handed boson \PZR decaying to three gluons and on an excited quark decaying via a vector boson to three quarks, as well as updated limits on a Kaluza--Klein gluon decaying via a radion to three gluons.
}
\hypersetup{
pdfauthor={CMS Collaboration},
pdftitle={Search for narrow trijet resonances in proton-proton collisions at sqrt{s}= 13 TeV},
pdfsubject={CMS},
pdfkeywords={CMS, LHC, BSM, trijet, exotica}} 
\maketitle 

The standard model (SM) of particle physics has several well-established shortcomings, notably lacking explanations for the nonzero neutrino masses, the issue of dark matter, the matter-antimatter asymmetry of the universe, the hierarchy problem, and the proliferation of quarks and leptons. 
These shortcomings have motivated a comprehensive program of searches for beyond-the-SM physics at high-energy colliders. 
Searches for new particles coupled to quarks and/or gluons are especially compelling at hadron colliders, such as the CERN LHC, since these particles could potentially be produced with significant cross sections. 
Searches for new particles decaying to two hadronic jets (dijets) have been performed numerous times at past and current colliders~\cite{Harris:2011bh,UA1,UA2-2,CDF4,D0-3,ATLAS13,ATLAS-boosted1,ATLAS12,CMS15,CMS-boosted5,CMS16}. 
Hadronic resonances with more complicated signatures have also been explored, including searches for the pair production of resonances decaying to two or more jets~\cite{CDF:2011xkb,PhysRevLett.111.031802,ATLAS:2017jnp,Sirunyan:2018rlj,CMS:2018pdq,ATLAS:2018umm,Sirunyan:2018duw,CMS-paired-dijet} and a search for resonances decaying to a jet and a boosted dijet resonance~\cite{CMS-trijet-dijet}.

In this Letter, we present the first generic search for singly produced resolved three-jet (trijet) resonances (\PX) using proton-proton ($\Pp\Pp$) collision data recorded by the CMS experiment from 2016 to 2018 at a center-of-mass energy of 13\TeV, corresponding to an integrated luminosity of 138\fbinv. 
Following techniques established by dijet resonance searches, the trijet invariant mass (\mjjj) spectrum is scrutinized for narrow excesses compatible with new resonances decaying to three hadronic jets.
No assumptions are made on the structure of the decay, except for the requirement that the final state contain three resolved jets with large transverse momentum (\pt). 
The SM background consists of events containing three or more jets produced exclusively through quantum chromodynamics (QCD), and is estimated from data by fitting the \mjjj spectrum with smoothly falling empirical functions.
The search for singly produced trijet resonances provides unique sensitivity for certain scenarios such as a spin-1 resonance dominantly produced via two off-shell gluons and decaying to three on-shell gluons. 
It also possesses complementary sensitivity to previous searches in scenarios such as a resonance decaying to another resonance (\PY) in association with a gluon or quark. 
Sensitivity to these representative scenarios is probed using three specific signal benchmark models that could potentially address the aforementioned SM shortcomings: a right-handed \PZ boson, \PZR, decaying directly to three gluons (\Pg)~\cite{Huitu:1996su}; a Kaluza--Klein excitation of a gluon, \PGKK, decaying via an intermediate radion to three gluons~\cite{Agashe:2016kfr,Agashe:2020wph}; and an excited quark, $\PQq^{*}$, decaying via an intermediate beyond-the-SM vector boson to three quarks (\PQq)~\cite{PhysRevD.42.815}. 
The latter two scenarios, which involve an initial resonance \PX and an intermediate resonance \PY, are referred to as cascade decays and are characterized by the mass ratio $\rhom=m_{\PY}/m_{\PX}$.
Values of \rhom between 0.2 and 0.8 that could lead to the resolved trijet signature are considered. 
At values of \rhom lower than 0.2, the \PY is produced with significant transverse momentum, and its decay products are reconstructed as a single jet; this scenario was explored in Ref.~\cite{CMS-trijet-dijet}. Supplemental information about the analysis is available in~\suppMaterial. Tabulated results are provided as a HEPData record for this analysis~\cite{hepdata}.

The CMS apparatus~\cite{CMS:2008xjf} is a multipurpose, nearly hermetic detector, designed to trigger on~\cite{CMS:2020cmk,CMS:2016ngn} and identify electrons, muons, photons, and charged and neutral hadrons~\cite{CMS:2020uim,CMS:2018rym,CMS:2014pgm}. 
The ``particle-flow'' algorithm~\cite{CMS:2017yfk} aims to reconstruct all individual particles in an event, combining information provided by the all-silicon inner tracker,  the crystal electromagnetic calorimeters (ECAL), and the brass-scintillator hadron calorimeters (HCAL), all operating inside a 3.8\unit{T} superconducting solenoid, with data from the gas-ionization muon detectors embedded in the flux-return yoke outside the solenoid. The reconstructed particles are used to build \PGt leptons, jets, and missing transverse momentum~\cite{CMS:2018jrd,CMS:2016lmd,CMS:2019ctu}.

Events of interest are selected using a two-tiered trigger system. The first level, composed of custom hardware processors, uses information from the calorimeters and muon detectors to select events at a rate of around 100\unit{kHz} with a fixed latency of about 4\mus~\cite{CMS:2020cmk}. The second level, known as the high-level trigger, consists of a farm of processors running a version of the full event reconstruction software optimized for fast processing, and reduces the event rate to around 1\unit{kHz} before data storage~\cite{CMS:2016ngn}.

The primary vertex is taken to be the vertex corresponding to the hardest scattering in the event, evaluated using tracking information alone~\cite{CMS-TDR-15-02}. The energy of photons is obtained from the ECAL measurement. The energy of electrons is determined from a combination of the electron momentum at the primary interaction vertex as determined by the tracker, the energy of the corresponding ECAL cluster, and the energy sum of all bremsstrahlung photons spatially compatible with originating from the electron track. The energy of muons is obtained from the curvature of the corresponding track. The energy of charged hadrons is determined from a combination of their momentum measured in the tracker and the matching ECAL and HCAL energy deposits, corrected for the response function of the calorimeters to hadronic showers. Finally, the energy of neutral hadrons is obtained from the corresponding corrected ECAL and HCAL energies.

For each event, hadronic jets are clustered from these reconstructed particles using the infrared- and collinear-safe anti-\kt algorithm~\cite{Cacciari:2008gp, Cacciari:2011ma} with a distance parameter of 0.4 (AK4). 
Jet momentum is determined as the vectorial sum of all particle momenta in the jet, and is found from simulation to be, on average, within 5\% to 10\% of the true momentum over the entire \pt spectrum and detector acceptance. 
Additional $\Pp\Pp$ interactions within the same or nearby bunch crossings (pileup) can contribute additional tracks and calorimetric energy depositions to the jet momentum. 
To mitigate this effect, charged particles identified to be originating from pileup vertices are discarded, and an offset correction is applied to correct for remaining contributions.
Jet energy corrections are derived from simulation to bring the measured response of jets to that of particle level jets on average. 
In situ measurements of the momentum balance in dijet, $\text{photon} + \text{jet}$, $\PZ + \text{jet}$, and multijet events are used to account for any residual differences in the jet energy scale between data and simulation~\cite{CMS:2016lmd}. 
The jet energy resolution amounts typically to 15\%--20\% at 30\GeV, 10\% at 100\GeV, and 5\% at 1\TeV~\cite{CMS:2016lmd}. 
Additional selection criteria are applied to each jet to remove jets potentially dominated by anomalous contributions from various subdetector components or reconstruction failures~\cite{CMS-PAS-JME-16-003}. 
Jets with $\pt<50\GeV$ are further required to pass a multivariate pileup jet rejection algorithm, which has 99\% efficiency for jets originating from the primary vertex of interest~\cite{CMS:2020ebo}.

To mitigate the loss of resolution from QCD radiation outside the jet boundaries, a ``wide--jet'' algorithm introduced for previous CMS dijet searches~\cite{CMS4} is used. 
Using the three leading jets with $\pt>100\GeV$ as seeds, the four-vectors of additional jets in the event with $\pt>30\GeV$ are added to the nearest seed jet within an optimized distance of $\Delta R\equiv\sqrt{(\Delta\eta)^2+(\Delta\phi)^2}<1.1$, where $\eta$ is the pseudorapidity and $\phi$ is the azimuthal angle in radians.  
The wide jet algorithm improves the sensitivity of the search to the signal cross section by $\sim$10\%-15\% (50\%-70\%) compared to the use of AK8 (AK4) jets. The trijet invariant mass, $\mjjj$, is defined as the invariant mass of the three wide jets obtained with this algorithm. 

Monte Carlo (MC) simulation is used to model the contributions from the production of new resonances that assume the Breit--Wigner line shape.
The $\PZR\to\Pg\Pg\Pg$ process is simulated at leading order (LO) in QCD with \PYTHIA~8.242~\cite{Sjostrand:2014zea}. 
Two width scenarios are simulated: the nominal left-right symmetric scenario from Ref.~\cite{Huitu:1996su} assuming SM-like couplings with a width of $\Gamma_{\PZR}/ m_{\PZR}\sim3\%$, and a narrow-width scenario with $\Gamma_{\PZR}/m_{\PZR}\sim0.01\%$ that represents the cases where the width is negligible with respect to the jet energy resolution. 
The $\PGKK$ process is modeled at LO with \MGvATNLO~2.7.3~\cite{madgraph}, interfaced with \PYTHIA for the parton shower.
The excited quark process is also modeled at LO in QCD with \PYTHIA. 
Simulated samples are also produced for the background QCD process, using \MGvATNLO~ 2.6.5 at LO with up to four partons at the matrix element level.
These background samples are used only for the development of the analysis strategy; the background estimation is performed using collision data. 
The NNPDF3.1~\cite{nnpdf31} parton distribution function (PDF) set at next-to-next-to-LO accuracy is used for all processes, along with the parton shower tune CP5~\cite{cp5}. 
The interaction of particles with the detector is simulated using the \GEANTfour toolkit~\cite{geant4}. 
The effects of pileup are incorporated by overlaying minimum bias events, simulated with \PYTHIA, on the hard scattering interactions, with the multiplicity distribution matching that observed in data. 

In the MC simulations, only signal events with a generated resonance mass greater than 85\% of the mass point under consideration are used. This requirement is found to have negligible impact on the search while helping to avoid uncertainties that are associated with the exact description of the far-off-shell low-mass tails.
The acceptance of this requirement, which depends on the width of the resonance, is denoted by $\mathcal{A}$.

Events are selected using a set of triggers requiring large hadronic activity, based on either the \pt of a single jet or the event \HT, defined as the scalar sum of jet \pt for all jets with $\pt>30\GeV$ and $\abs{\eta}<3.0$. 
The typical thresholds deployed for \HT triggers were 900 (1050)\GeV and 360 (500)\GeV for the \pt triggers in 2016 (2017--2018).
To avoid the phase space where the trigger is inefficient, we apply a selection of $\mjjj>1.50\TeV$ (2016) or $\mjjj>1.76\TeV$ (2017--2018). 
The trigger and \mjjj thresholds for 2016 data are lower because of the lower instantaneous luminosity.
The fully reconstructed events are required to have at least three reconstructed jets with $\pt>100\GeV$ and $\abs{\eta}<2.5$, prior to the application of the wide--jet algorithm. 
The aforementioned jet identification criteria, which have an efficiency of more than 98\%--99\%, are applied to each jet used to construct the wide jet.
Since the QCD background tends to have jets with higher values of $\abs{\eta}$ than the signal processes, the sensitivity of the analysis is improved by requiring that the maximum difference in $\eta$ between any two of the three wide jets, $\Delta\eta_{\mathrm{max}}$, be less than 1.6. 
Similarly, the maximum difference in $\Delta R$ between any two wide jets must be less than 3.0. 
The efficiencies of the above selection requirements are calculated in simulation to be ${\sim}30\%$ for the \PZR process and between 23\%--35\%~(26\%--37\%) for the \PGKK ($\PQq^{*}$) process for mass values between 2.0 and 9.0\TeV. 
For resonance masses below ${\sim}2\TeV$, the efficiencies decrease because the signal peaks are truncated by the lower \mjjj threshold, reaching ${\sim}20\%$ for the \PZR process, ${\sim}15\%$ for the \PGKK process, and ${\sim}25\%$ for the $\PQq^{*}$ process at a mass of 1.75\TeV.

\begin{figure}
    \centering
    \includegraphics[width=0.49\textwidth]{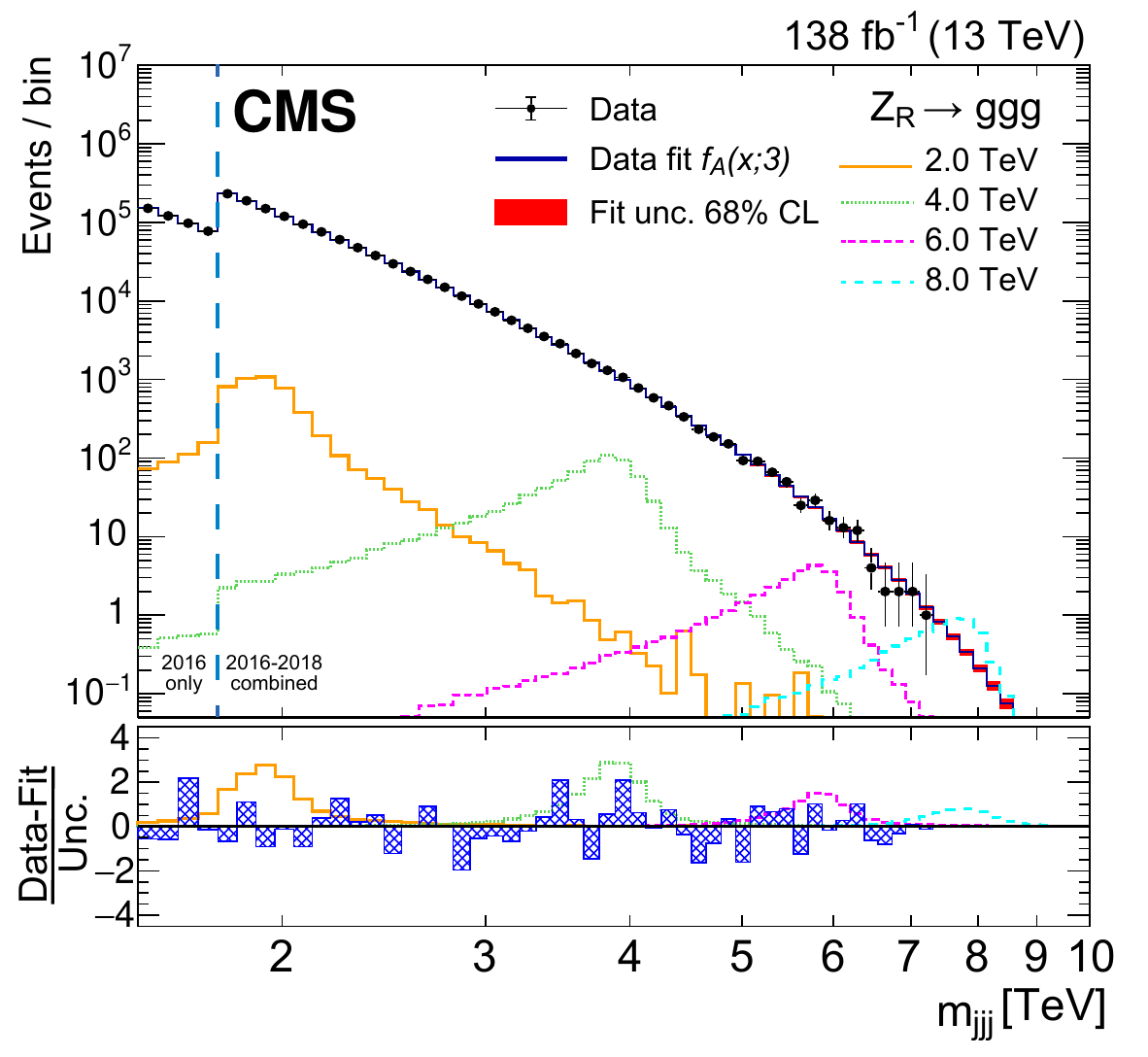}
    \caption{
The observed $\mjjj$ distribution and the background-only fit to the data using the $f_A$ fit function.
Uncertainties in the fit that correspond to the 68\% confidence level are depicted with the red band.
The expected $\mjjj$ distributions for $\PZR$ signal masses of 2.0, 4.0, 6.0, and 8.0\TeV, with nominal width of ${\sim} 3\%$, are also shown.
For illustration purposes, the normalizations correspond to $\sigma\mathcal{B}$ values of 200, 50, 20, and 20\fb, respectively. 
Only 2016 data are shown for $\mjjj<1.76\TeV$ because of the higher trigger thresholds in 2017 and 2018.
In the bottom panel, the blue hatched bars show the difference between the observed data and the background prediction divided by the statistical uncertainty, along with expectations for the example $\PZR$ signal points.}
    \label{fig:mjjj}
\end{figure}

The background, dominantly arising from the QCD multijet process, is modeled by fitting the \mjjj distribution with smoothly falling, empirical functions that have been used in previous searches~\cite{CMS-trijet-dijet,CMS-paired-dijet}.
Three families of functions are considered:
\begin{equation}
    \begin{aligned}
    f_A(x;\,N) &= p_{0} \frac{(1-x)^{p_{1}}}{x^{\sum_{i=2}^{N} p_{i}\log^{i-2}(x)}}, \\
    f_B(x;\,N) &= p_{0} \frac{\re^{-p_{1}x}} {x ^ {\sum_{i=2}^{N}p_{i}\log^{i-2}(x)}}, \\
    f_C(x;\,N) &= p_{0} x^{\sum_{i=1}^{N} p_{i}\log^{i-1}(x)},
    \end{aligned}
\end{equation}
where $x=\mjjj/\sqrt{s}$, $p_{i}~(i = 0, 1, \ldots, N)$ are free parameters, and $N$ is the order of the fit function, which is determined using a Fisher F-test~\cite{ftest}. 
It is found that the third order yields the best fits to the data for all three families of functions. The p-values obtained from a generalized $\chi^{2}$ goodness-of-fit test~\cite{Cousins:2018tiz} yield 0.62, 0.53, and 0.34 for the $f_A$, $f_B$, and $f_C$ functions, respectively.
The signal is extracted using a binned maximum-likelihood fit based on a profile likelihood ratio test statistic~\cite{LHCCLs}, with variable bin size approximating the $\mjjj$ resolution. 
The signal component of the fit is taken from binned templates derived from simulation, while the background component corresponds to the integral of the fit function across each bin.
The three fit functions are incorporated into a single likelihood using a discrete nuisance parameter~\cite{Dauncey:2014xga}. This parameter is profiled in an analogous way to continuous nuisance parameters to consider the extra uncertainty brought by the various possible choices of the background functional form. 
The background parameters are included as freely floating nuisance parameters. 
The three data-taking years are treated separately in the fit, in particular, with independent background parametrizations for each year, and are combined statistically into a single likelihood. 
The combined result of the background-only fits with the function $f_A(x;\,3)$ is shown in Fig.~\ref{fig:mjjj} as an example. 

To gauge the systematic biases in the signal extraction, bias tests are performed using pseudo-data produced from the background-only fits with injected signal events. 
The bias is found to be negligible.
Systematic uncertainties in the signal are incorporated in the likelihood as constrained nuisance parameters. 
The jet energy resolution and scale uncertainties translate to ${\approx}10\%$ and ${\approx}1\%$ uncertainties in the width and the position of the trijet mass shape for the signal, respectively.
The integrated luminosities for the 2016, 2017, and 2018 data-taking years have 1.2\%--2.5\% individual uncertainties~\cite{CMS-LUM-17-003,CMS-PAS-LUM-17-004,CMS-PAS-LUM-18-002}, while the overall uncertainty for the 2016--2018 period is 1.6\%. 
Finally, the uncertainty in the predicted signal yield that stems from the pileup reweighting procedure is ${\approx}3\%$ for the three years, and is uncorrelated between years. 
The most significant uncertainties are those associated with background fit parameters in the background prediction, which are evaluated to have an impact on the fitted signal strength of ${\approx}10\%$ over the entire search range.

No significant excesses compatible with the signal hypotheses are observed.
For the $\PX\to\Pg\Pg\Pg$ signal scenario, the largest deviation is observed at $m_{\PX}=4.1\TeV$, corresponding to local significance values of 2.1 and 2.2 standard deviations for the nominal and narrow width hypotheses, respectively. 
Taking into account the look-elsewhere effect~\cite{Gross:2010qma}, the global significance values are 0.2 and 0.3 standard deviations, respectively. 
For the $\PX\to\PY(\Pg\Pg)\Pg$ cascade decay scenario, the largest deviation is observed for $\rhom=0.3$ and $m_{\PX}=4.1\TeV$, with a local (global) significance of 2.2 (0.4) standard deviations. 
For the $\PX\to\PY(\Pq\Pq)\Pq$ cascade decay scenario, the largest deviation is observed at $\rhom=0.7$ and $m_{\PX}=3.9\TeV$, and corresponds to a local (global) significance of 2.1 (0.3) standard deviations. 

\begin{figure}
    \centering
    \includegraphics[width=0.49\textwidth]{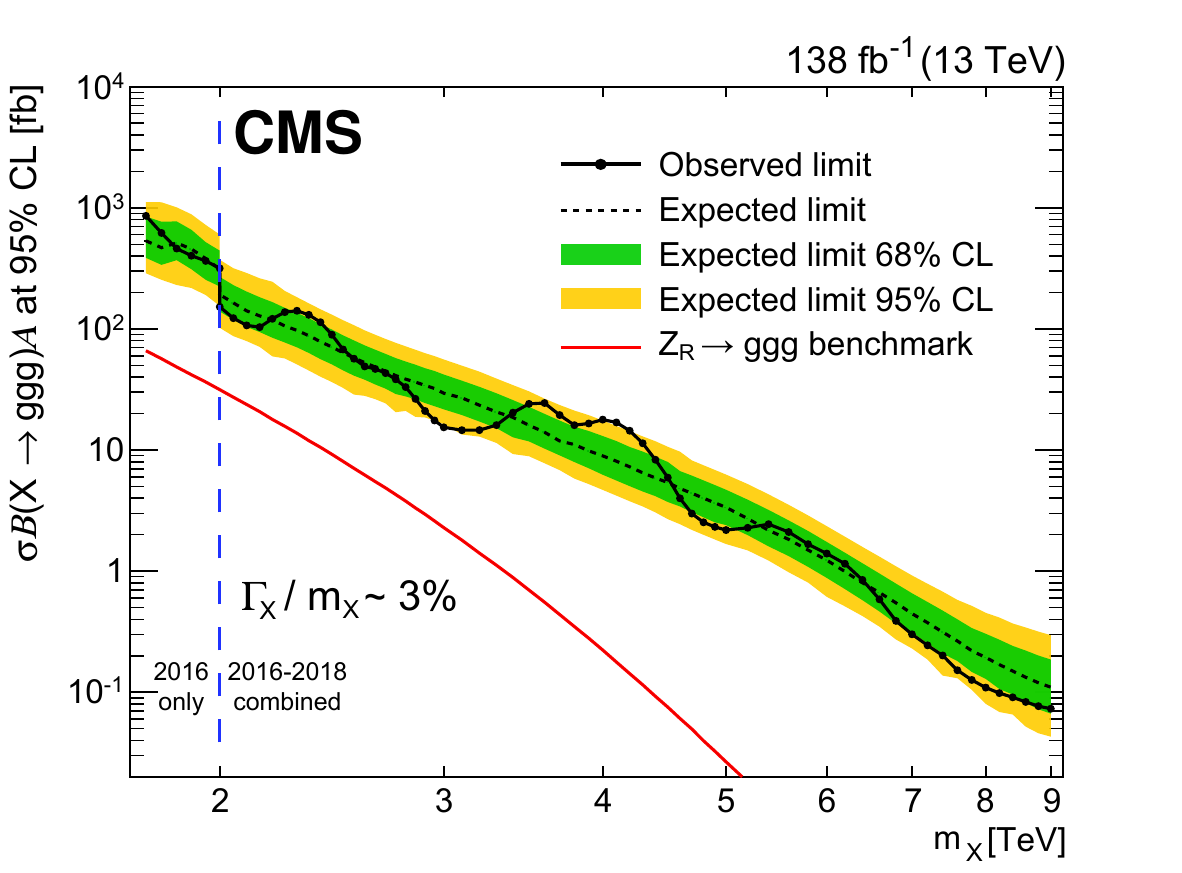}
    \includegraphics[width=0.49\textwidth]{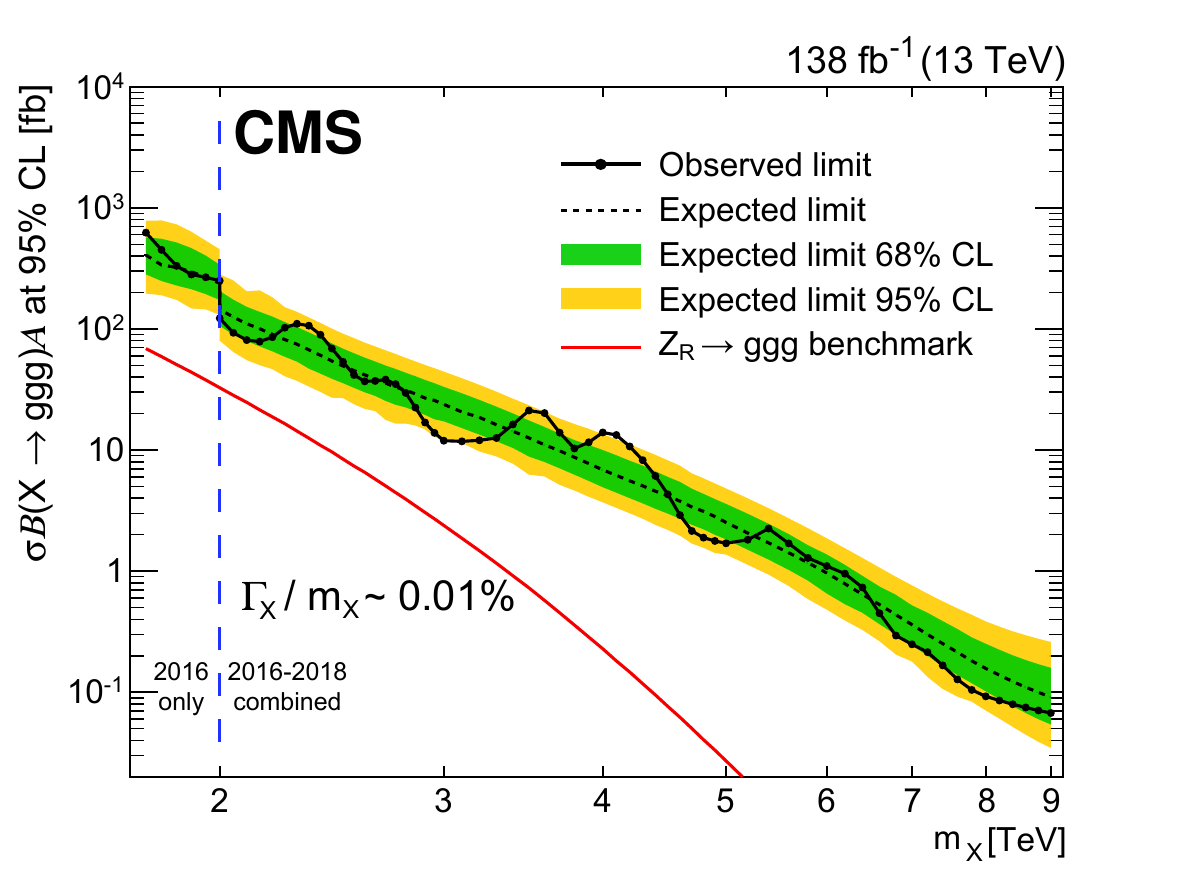}
    \caption{
      Limits at 95\% \CL on $\sigma\mathcal{B}(\PX\to\Pg\Pg\Pg)\mathcal{A}$ for the nominal (\cmsLeft) and narrow-width (\cmsRight) scenarios. 
      Only 2016 data are used to derive limits below 2.0\TeV because of the higher trigger thresholds in 2017 and 2018. 
      Theoretical predictions assuming SM-like couplings are depicted with red curves.}
    \label{fig:limits_ZR}
\end{figure}

\begin{figure}
    \centering
    \includegraphics[width=0.49\textwidth]{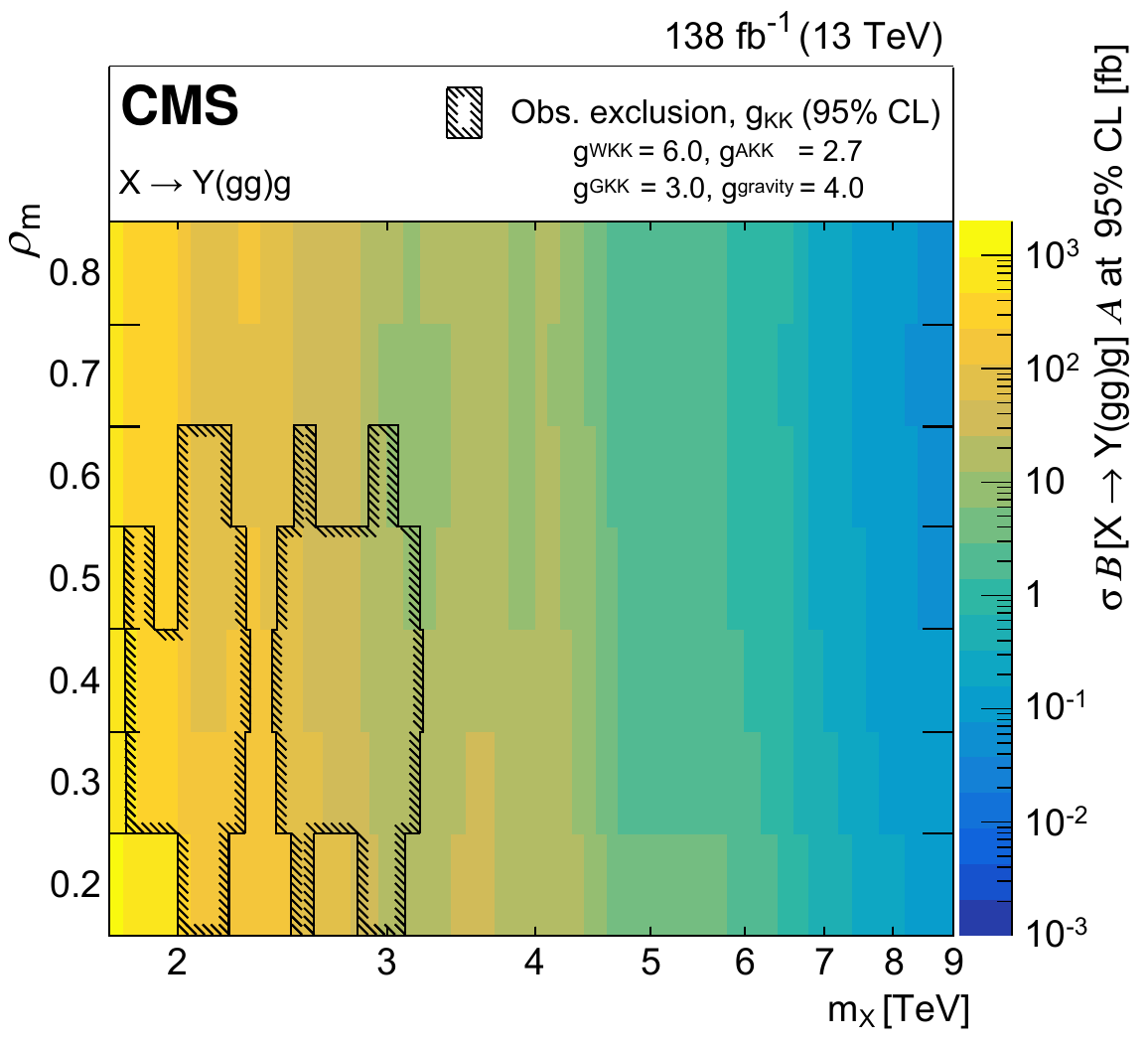}
    \hfill
    \includegraphics[width=0.49\textwidth]{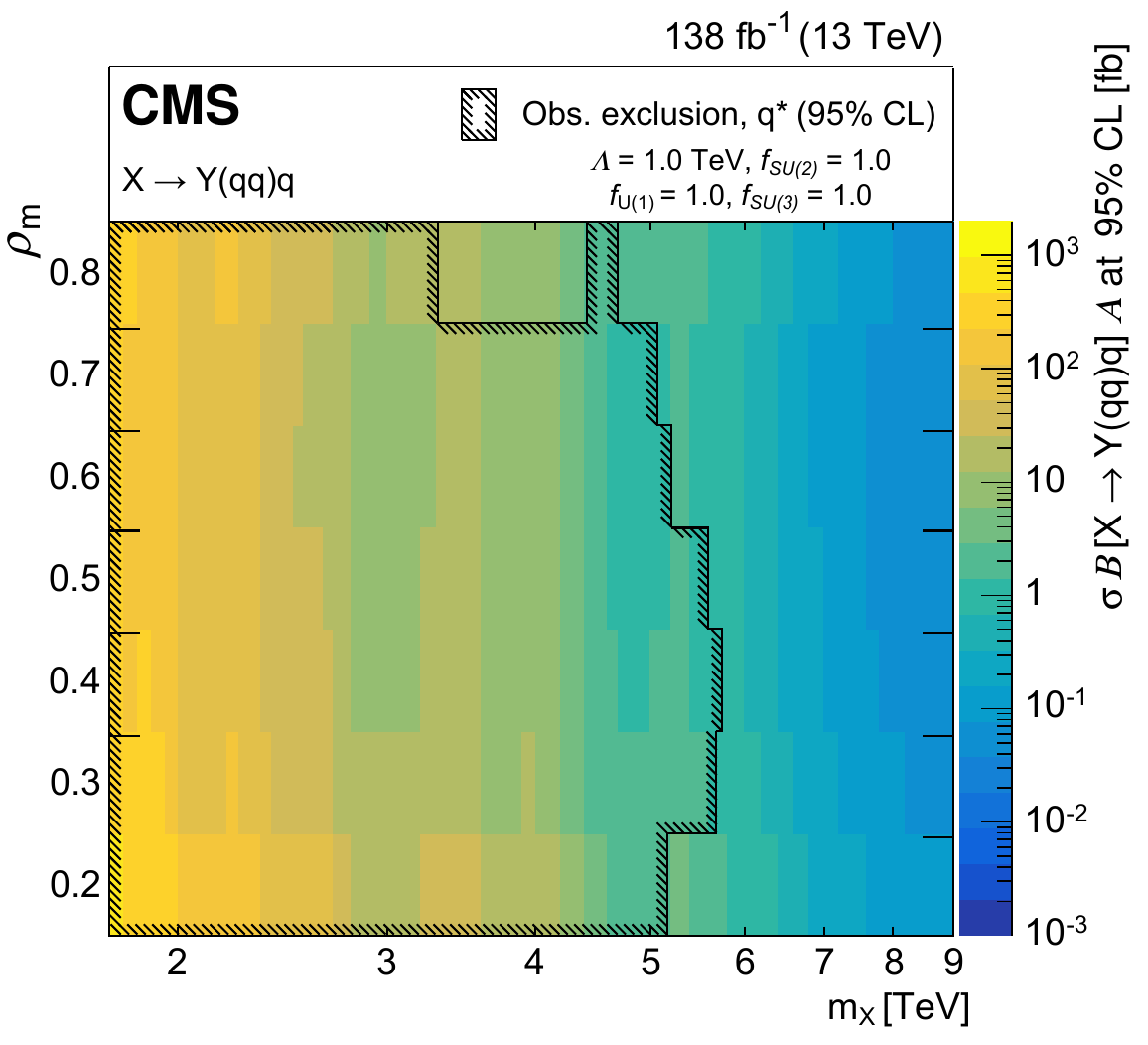}
    \caption{
      Observed limits at 95\% \CL as a function of $m_{\PX}$ and $\rhom$ on $\sigma\mathcal{B}[\PX\to\PY(\Pg\Pg)\Pg]\mathcal{A}$ (\cmsLeft) and $\sigma\mathcal{B}[\PX\to\PY(\PQq\PQq)\PQq]\mathcal{A}$ (\cmsRight). 
      Only 2016 data are used to derive limits below 2.0\TeV because of the higher trigger thresholds in 2017 and 2018. 
      The legend shows the model parameters that are defined in ~\cite{Agashe:2020wph,PhysRevD.42.815} for the chosen benchmark, and their corresponding mass exclusion ranges are depicted with areas inside the black hatched contours.}
    \label{fig:limits_cascade}
\end{figure}

Limits are set for the first time at 95\% confidence level (\CL) on the product of the signal process production cross section, branching fraction, and acceptance ($\sigma\mathcal{B}\mathcal{A}$) to three resolved jets using the asymptotic \CLs procedure~\cite{CLS1,CLS2,Cowan:2010js} in the mass range between 1.75\TeV and 9.00\TeV. 
Only 2016 data are used to derive limits below 2.0\TeV because of the higher trigger thresholds deployed in 2017 and 2018.
Limits are shown in Fig.~\ref{fig:limits_ZR} for the three-body decay of new resonances, ranging from 0.073 (0.067)\fb to 0.86 (0.62)\pb for the nominal (narrow) width hypotheses. 
Theoretical predictions of $\sigma\mathcal{B}\mathcal{A}$ assuming SM-like couplings are depicted with red curves. 
The current data set does not provide sufficient sensitivity to constrain the $\PZR$ model.
Limits on the cascade decay of new resonances into $\Pg\Pg\Pg$ ($\PQq\PQq\PQq$) are presented in Fig.~\ref{fig:limits_cascade}, which range from 0.061 (0.047)\fb to 1.2 (0.49)\pb. 
Benchmark model coupling parameters~\cite{Agashe:2020wph,PhysRevD.42.815} are shown in the legend, and their corresponding mass exclusion ranges are depicted with areas inside the black hatched contours. 
At $\rhom = 0.2$, this search achieves a similar level of sensitivity to that of the previous search~\cite{CMS-trijet-dijet} for the $\Pg\Pg\Pg$ decay mode.

In summary, the first generic search for singly produced new particles decaying to three hadronic jets has been presented. 
The search uses proton-proton collision data at $\sqrt{s}=13\TeV$ recorded by the CMS experiment in 2016--2018, corresponding to an integrated luminosity of $138\fbinv$. 
The three-jet invariant mass spectrum is scanned for narrow peaks corresponding to new particles. 
No significant excesses above the standard model background expectations are observed.
Limits are set on the product of the production cross section, branching fraction, and acceptance to three resolved jets. 
The results are interpreted in the context of a new right-handed boson \PZR decaying to three gluons, a Kaluza--Klein gluon \PGKK decaying via an intermediate radion to three gluons ($\Pg\Pg\Pg$), and an excited quark decaying via a vector boson to three quarks ($\PQq\PQq\PQq$). 
This is the first search for the three-body decay of singly produced high-mass resonances \PX into three resolved jets at the LHC, and also the first search for \PX that decays into three resolved jets through an intermediate resonance \PY with a mass ratio $m_{\PY}/m_{\PX}$ within 0.3--0.8 for the $\Pg\Pg\Pg$ decay mode and 0.2--0.8 for the $\PQq\PQq\PQq$ decay mode, significantly extending the model parameter space explored by a previous search~\cite{CMS-trijet-dijet}.

\begin{acknowledgments}
We congratulate our colleagues in the CERN accelerator departments for the excellent performance of the LHC and thank the technical and administrative staffs at CERN and at other CMS institutes for their contributions to the success of the CMS effort. In addition, we gratefully acknowledge the computing centers and personnel of the Worldwide LHC Computing Grid and other centers for delivering so effectively the computing infrastructure essential to our analyses. Finally, we acknowledge the enduring support for the construction and operation of the LHC, the CMS detector, and the supporting computing infrastructure provided by the following funding agencies: SC (Armenia), BMBWF and FWF (Austria); FNRS and FWO (Belgium); CNPq, CAPES, FAPERJ, FAPERGS, and FAPESP (Brazil); MES and BNSF (Bulgaria); CERN; CAS, MoST, and NSFC (China); MINCIENCIAS (Colombia); MSES and CSF (Croatia); RIF (Cyprus); SENESCYT (Ecuador); MoER, ERC PUT and ERDF (Estonia); Academy of Finland, MEC, and HIP (Finland); CEA and CNRS/IN2P3 (France); SRNSF (Georgia); BMBF, DFG, and HGF (Germany); GSRI (Greece); NKFIH (Hungary); DAE and DST (India); IPM (Iran); SFI (Ireland); INFN (Italy); MSIP and NRF (Republic of Korea); MES (Latvia); LAS (Lithuania); MOE and UM (Malaysia); BUAP, CINVESTAV, CONACYT, LNS, SEP, and UASLP-FAI (Mexico); MOS (Montenegro); MBIE (New Zealand); PAEC (Pakistan); MES and NSC (Poland); FCT (Portugal); MESTD (Serbia); MCIN/AEI and PCTI (Spain); MOSTR (Sri Lanka); Swiss Funding Agencies (Switzerland); MST (Taipei); MHESI and NSTDA (Thailand); TUBITAK and TENMAK (Turkey); NASU (Ukraine); STFC (United Kingdom); DOE and NSF (USA).    
\end{acknowledgments}

\bibliography{auto_generated}

\ifthenelse{\boolean{cms@external}}{}{
\clearpage

\appendix
\numberwithin{table}{section}
\section{Supplemental material\label{app:suppMat}}
\begin{figure*}[htpb!]
    \centering
    \includegraphics[width=0.48\textwidth]{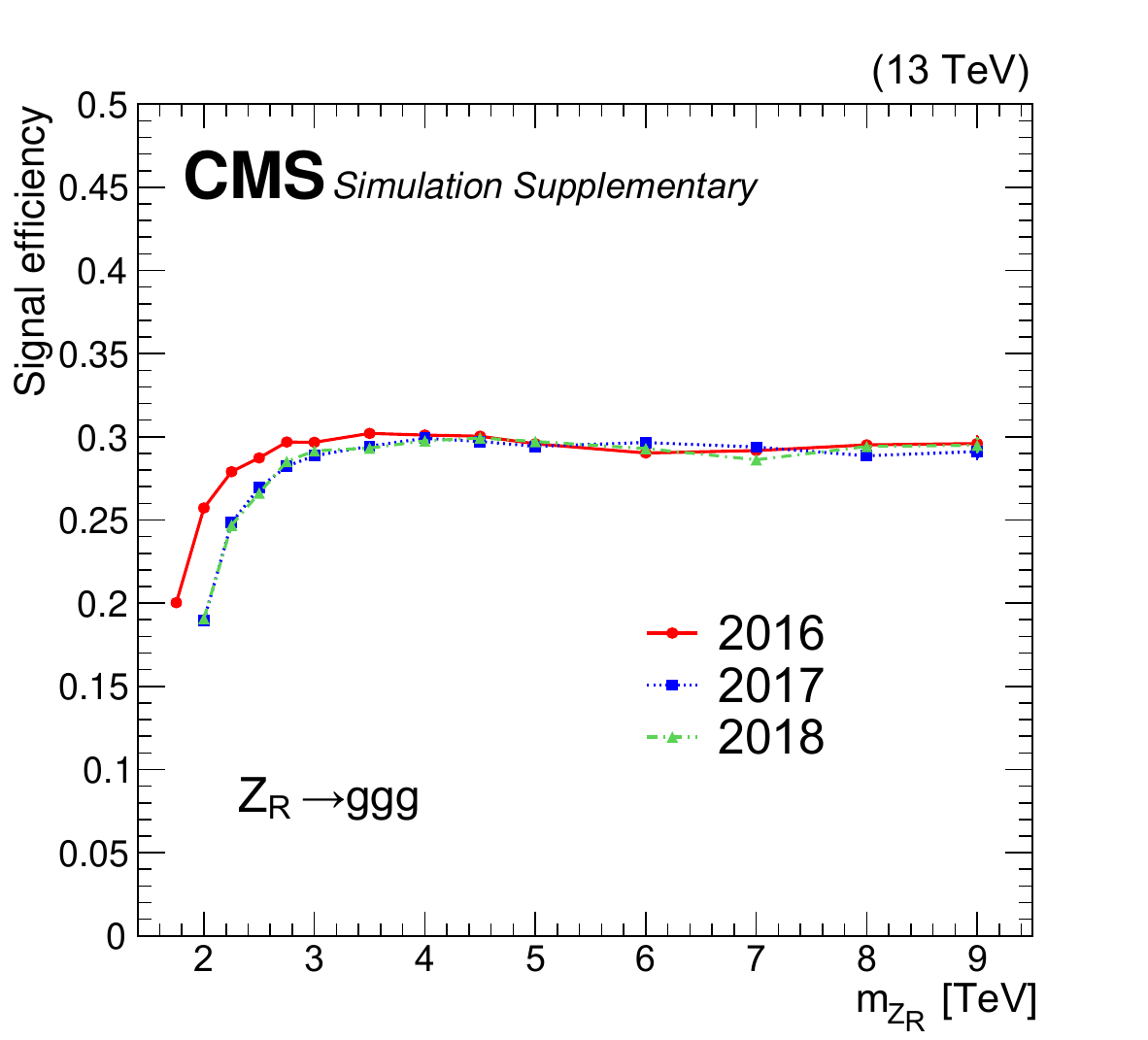}
    \includegraphics[width=0.48\textwidth]{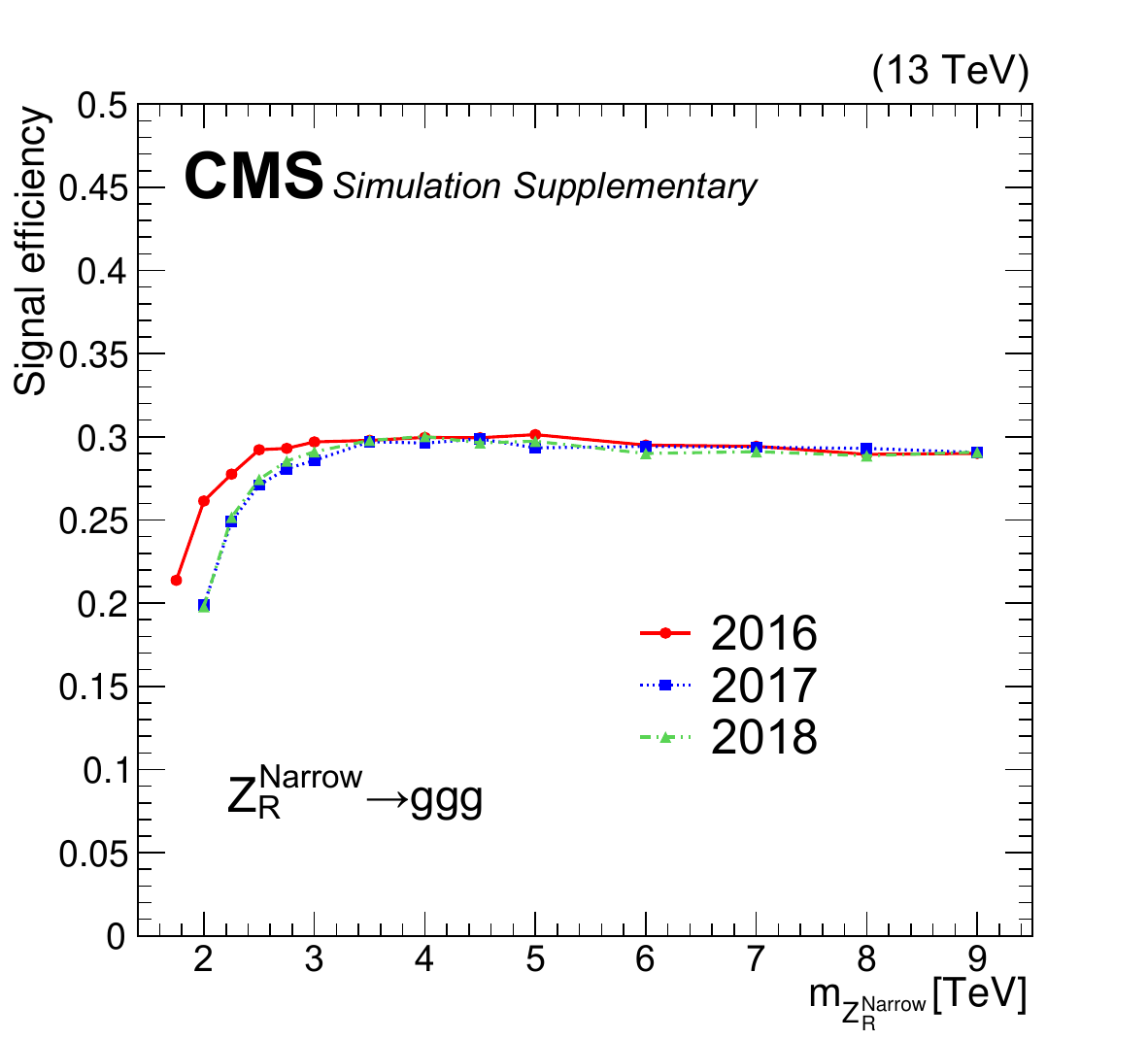}
    \includegraphics[width=0.48\textwidth]{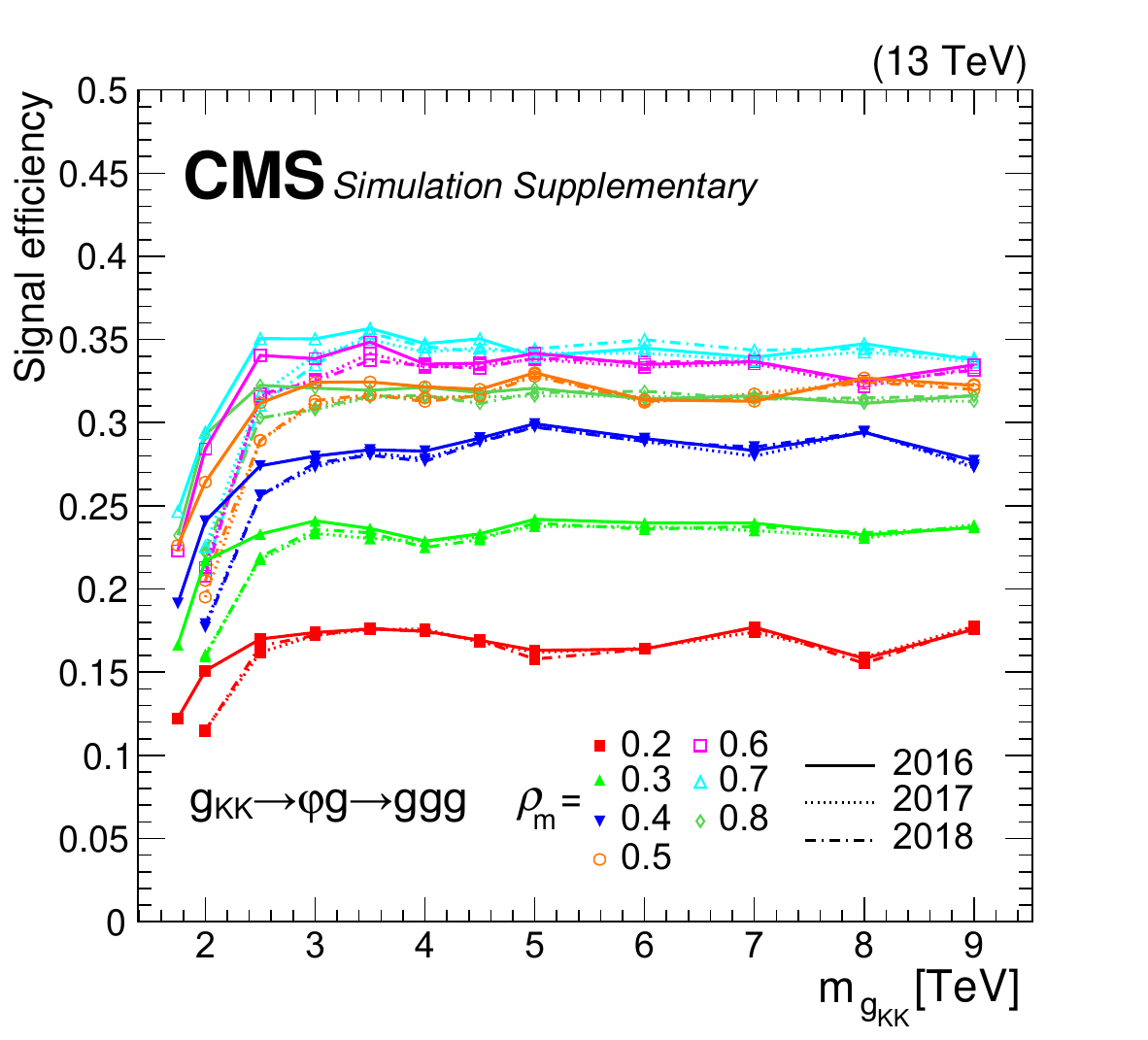}
    \includegraphics[width=0.48\textwidth]{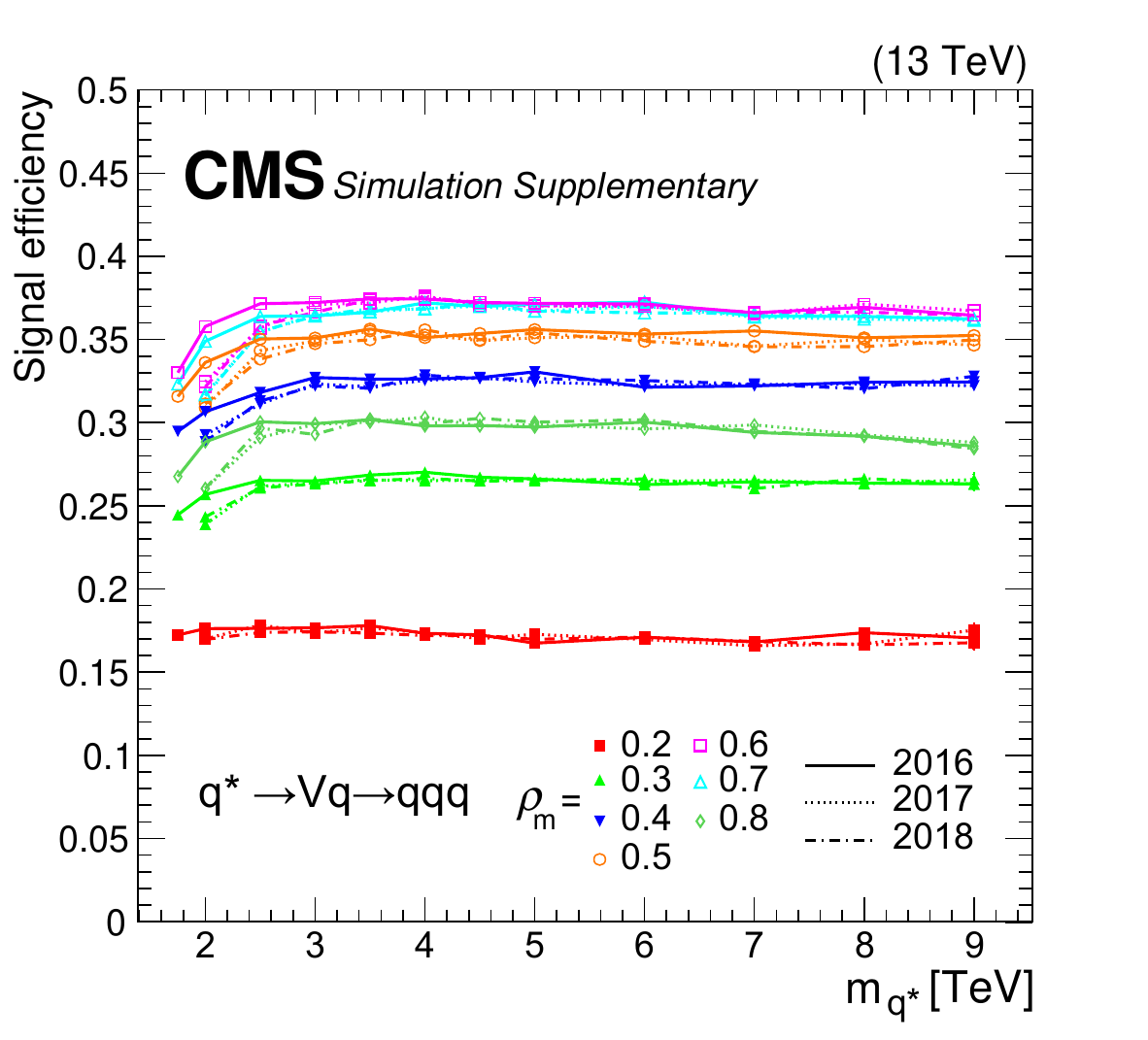}
    \caption{
        Efficiencies of the selection requirements on the benchmark signal processes: $\PZR\to\Pg\Pg\Pg$ with nominal width $\Gamma_{\PZR}/ m_{\PZR}\sim3\%$ (top left), $\PZR\to\Pg\Pg\Pg$ with narrow width $\Gamma_{\PZR}/ m_{\PZR}\sim0.01\%$ (top right), $\PGKK\to\varphi(\Pg\Pg)\Pg$ (bottom left), and $\PQq^{*}\to \text{V}(\PQq\PQq)\PQq$ (bottom right). $\varphi$ is the radion and $\text{V}$ is a beyond-the-SM vector boson.
        The efficiencies for 2016, 2017, and 2018 data-taking conditions are shown separately. 
        The bottom two figures also show efficiencies for different $\rho_{m}$ scenarios for cascade decays with intermediate resonances. 
    }
    \label{fig:efficiency}
\end{figure*}

\begin{figure*}[htpb!]
    \centering
    \includegraphics[width=0.48\textwidth]{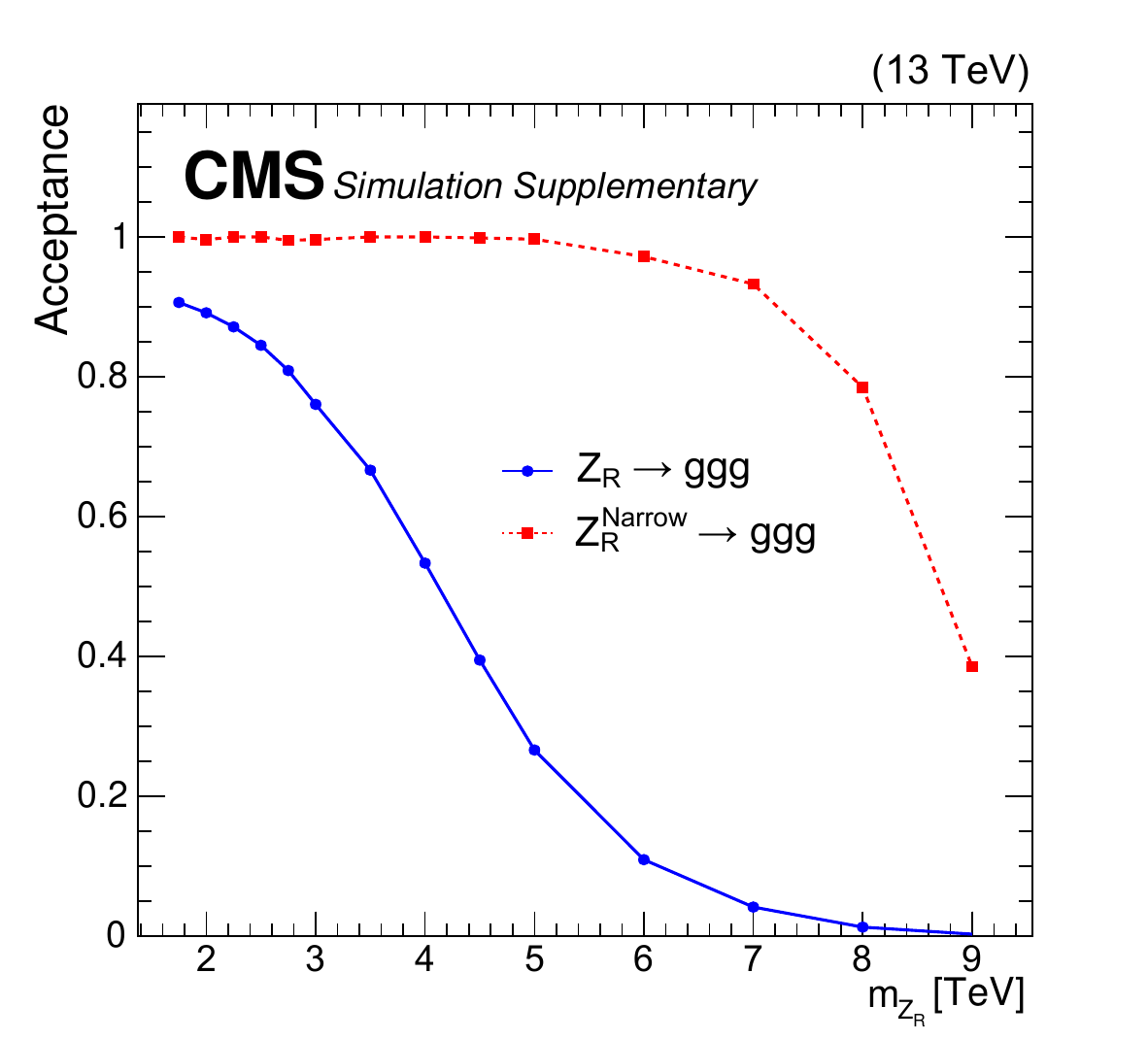}
	  \includegraphics[width=0.48\textwidth]{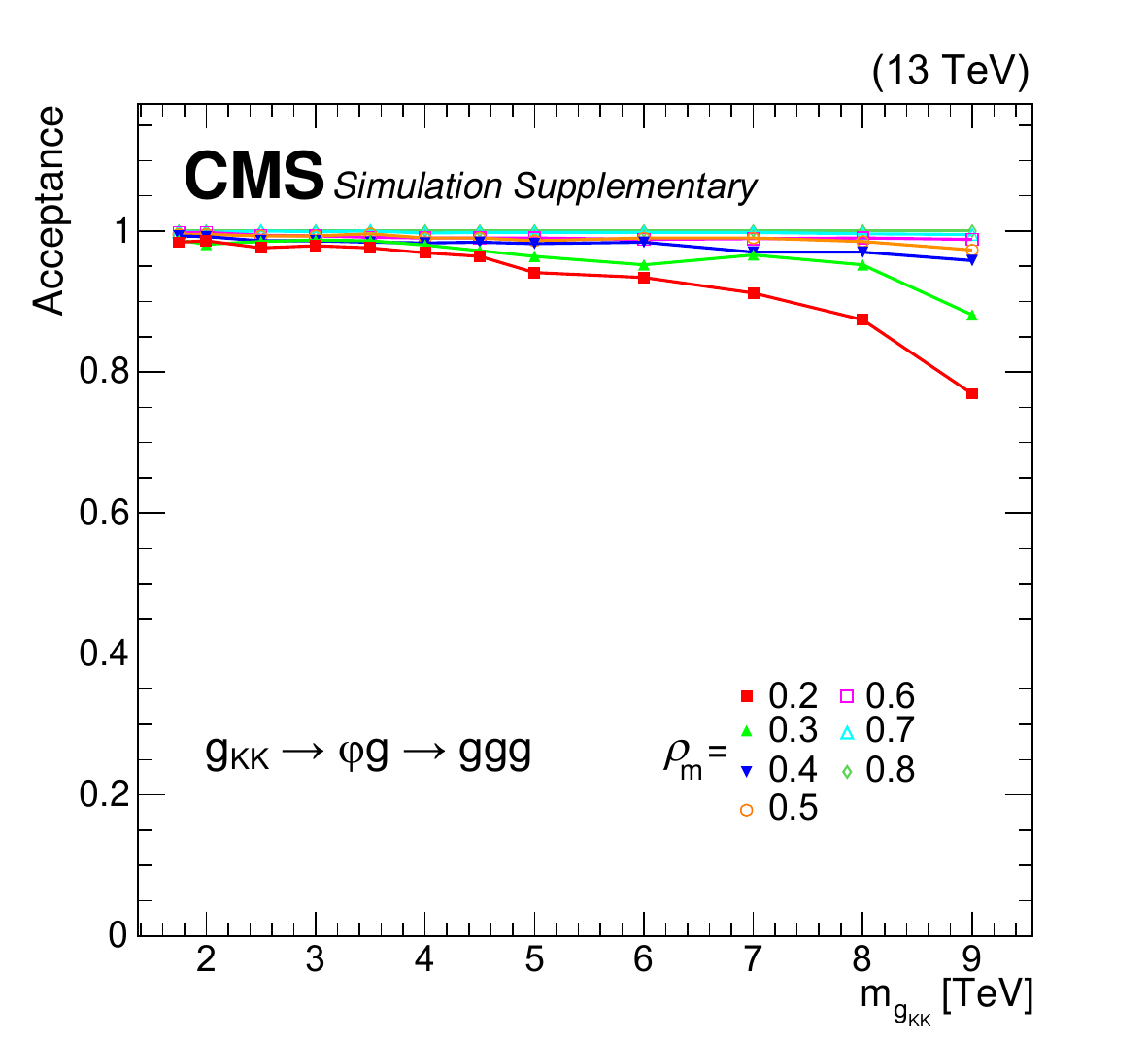}
	  \includegraphics[width=0.48\textwidth]{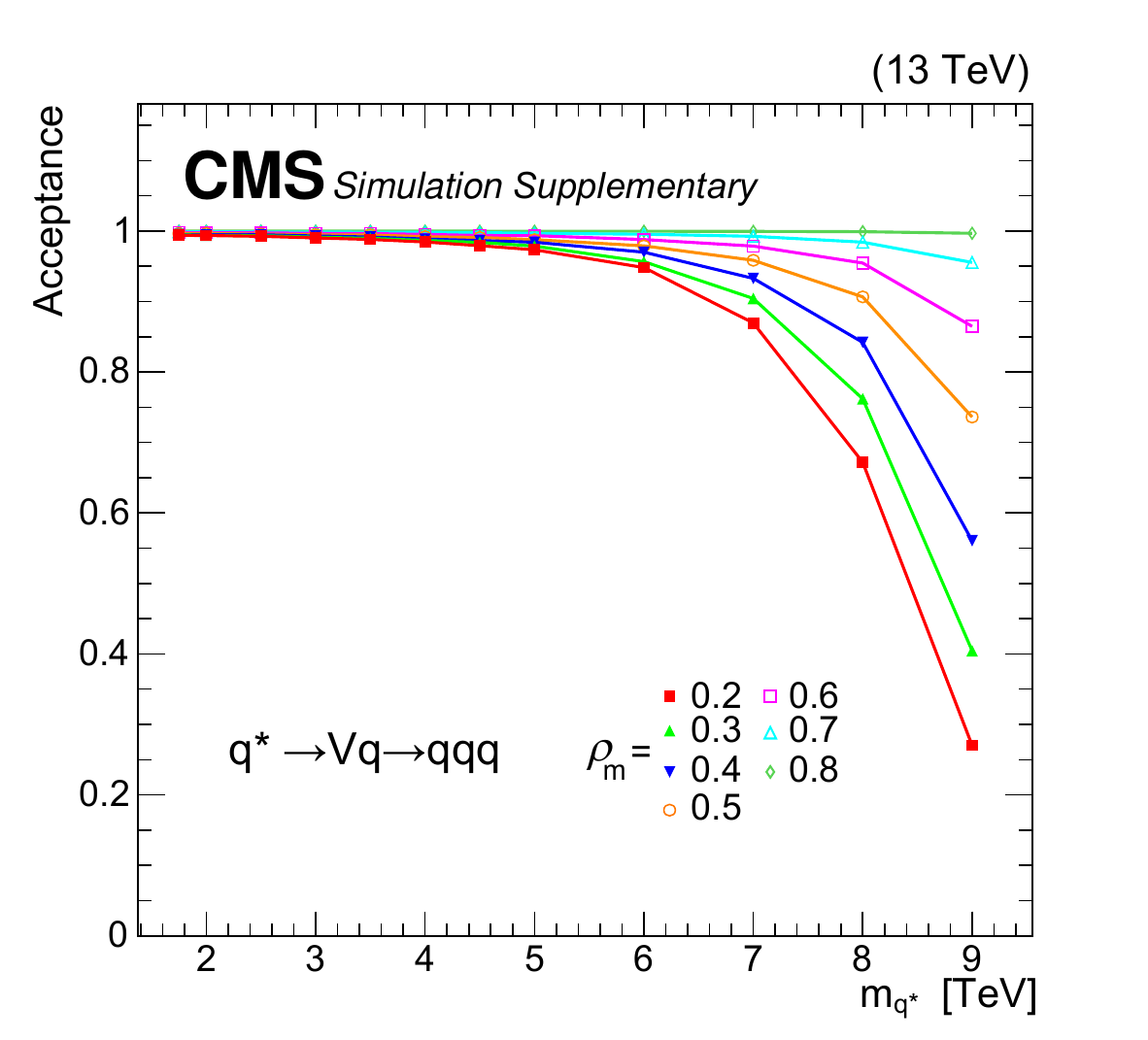}
    \caption{Acceptance of the signal selection requirement $m_{\PX}^{GEN}/m_{\PX}^{\text{input}}>85\%$ on the benchmark signal process $\PZR\to\Pg\Pg\Pg$ (top), $\PGKK\to\varphi(\Pg\Pg)\Pg$ (bottom left), and $\PQq^{*}\to \text{V}(\PQq\PQq)\PQq$ (bottom right). $m_{\PX}^{\text{GEN}}$ is the mass of the new resonances generated by the MC simulation. $m_{\PX}^{\text{input}}$ is the resonance mass point under consideration. The acceptance is defined as $\mathcal{A} = N$(events with $m_{\PX}^{\text{GEN}}/m_{\PX}^{\text{input}}>85\%$) / $N$(events generated in the full phase space defined by the CMS default generator settings).}
    \label{fig:acceptance}
\end{figure*}

\begin{figure*}[htpb!]
    \centering
    \includegraphics[width=0.48\textwidth]{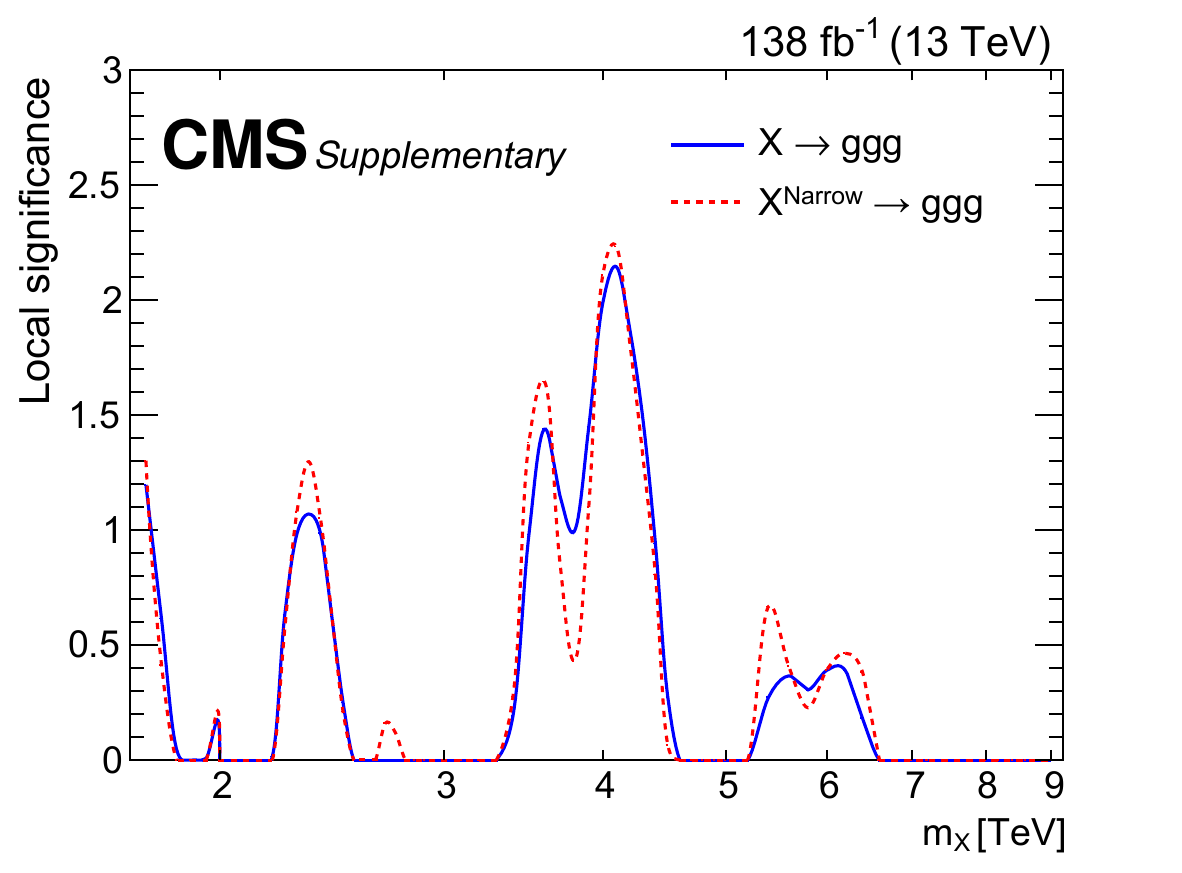}
    \caption{The observed local significance for a $\Pg\Pg\Pg$ resonance versus $\PX$ mass, shown for resonances with nominal width (blue) and narrow width (red). The most significant excesses correspond to 2.1 (2.2) standard deviations.}
    \label{fig:significance_ZR}
\end{figure*}

\begin{figure*}[htpb!]
    \centering
    \includegraphics[width=0.48\textwidth]{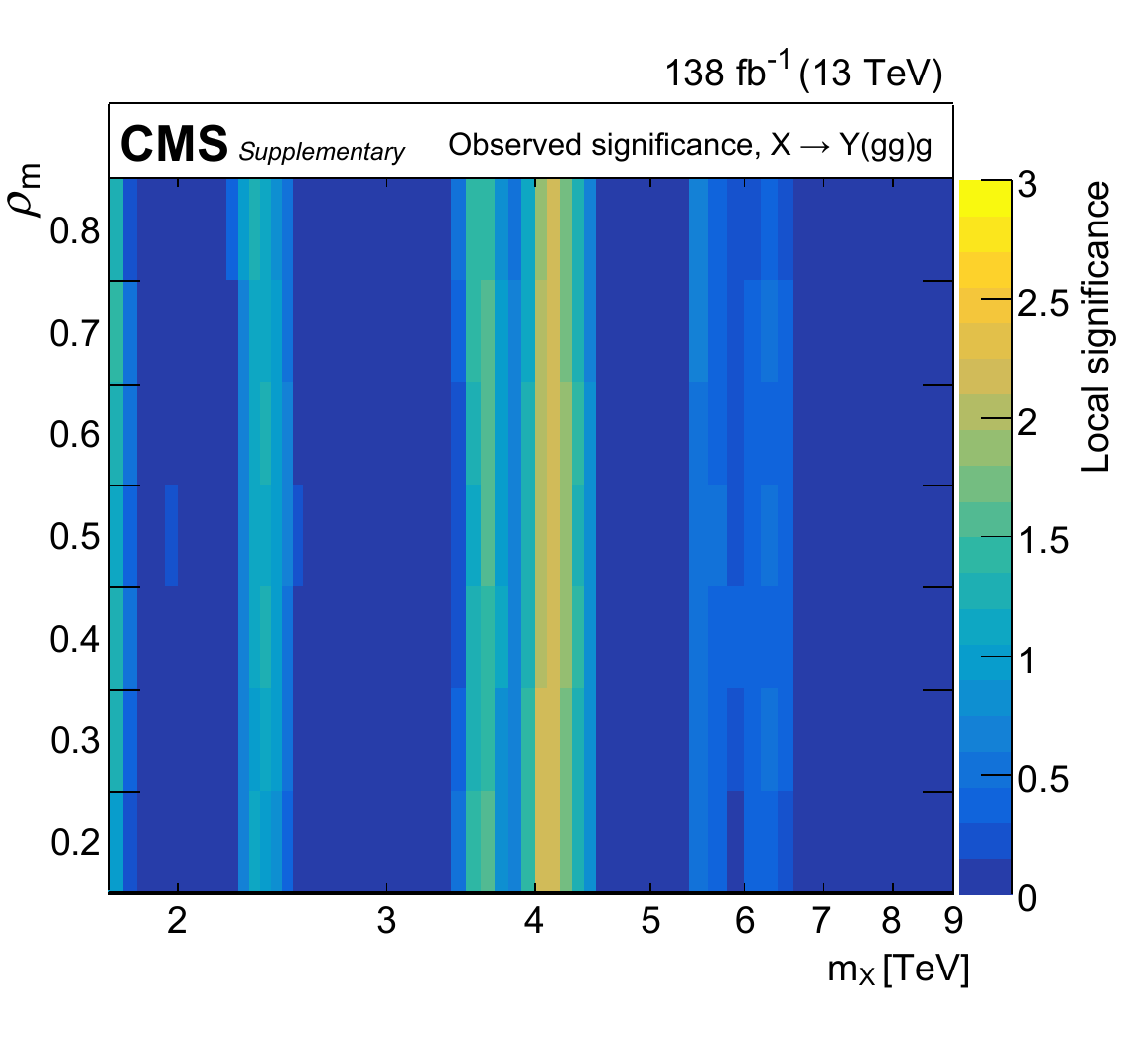}
    \includegraphics[width=0.48\textwidth]{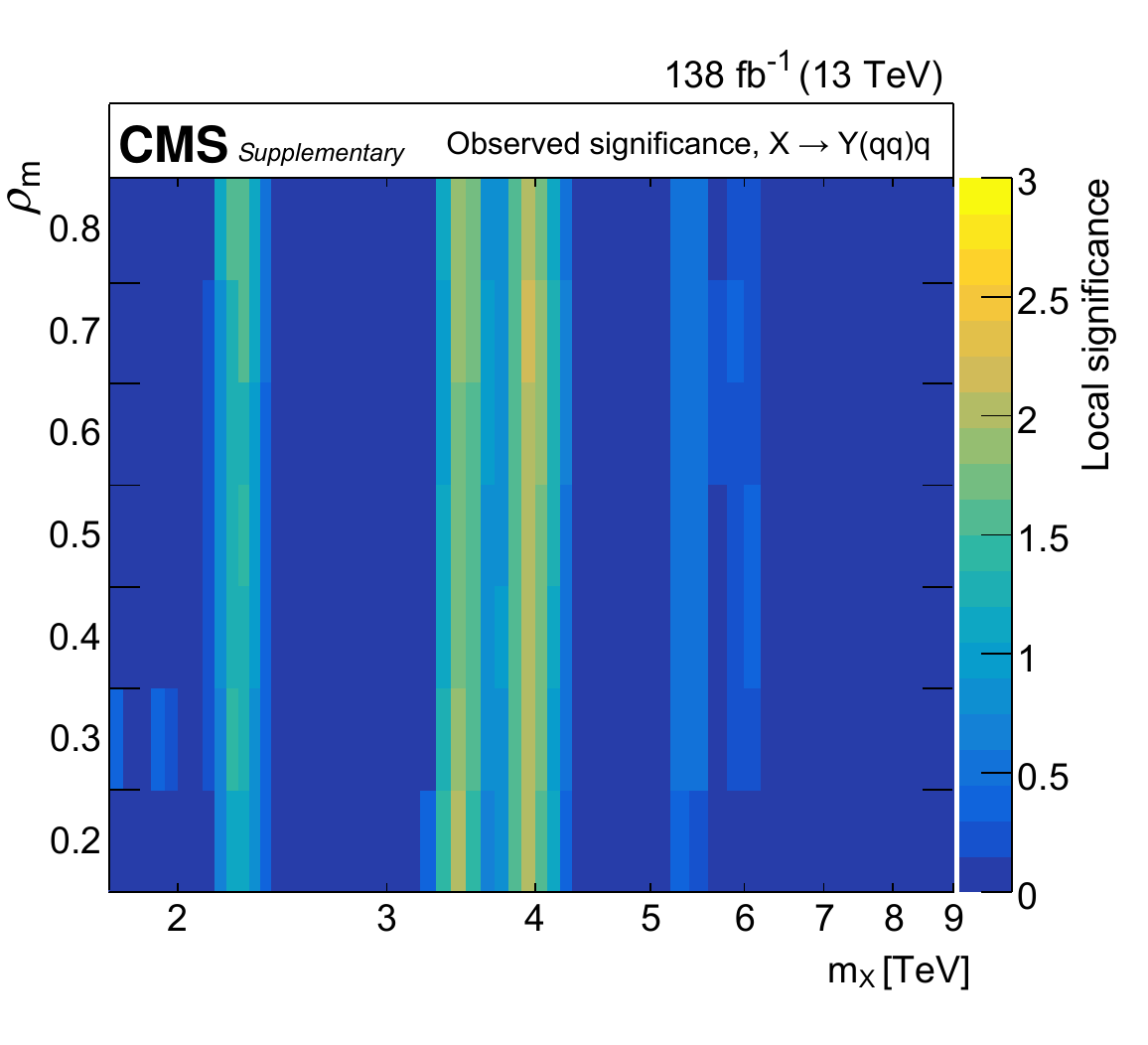}
    \caption{The observed local significance versus $m_{\PX}$ and $\rho_{m}$ for resonances decaying via a cascade. The largest deviations are observed at $\rho_{m} = 0.3, m_{\PX} = 4.1\TeV$ for $\PX\to\PY(\Pg\Pg)\Pg$ and $\rho_{m} = 0.7, m_{\PX} = 3.9\TeV$ for $\PX\to\PY(\PQq\PQq)\PQq$. The corresponding local (global) significance values are 2.2 (0.4) and 2.1 (0.3) standard deviations, respectively.}
    \label{fig:significance_cascade}
\end{figure*}

\begin{figure*}[htpb!]
    \centering
    \includegraphics[width=0.4\textwidth]{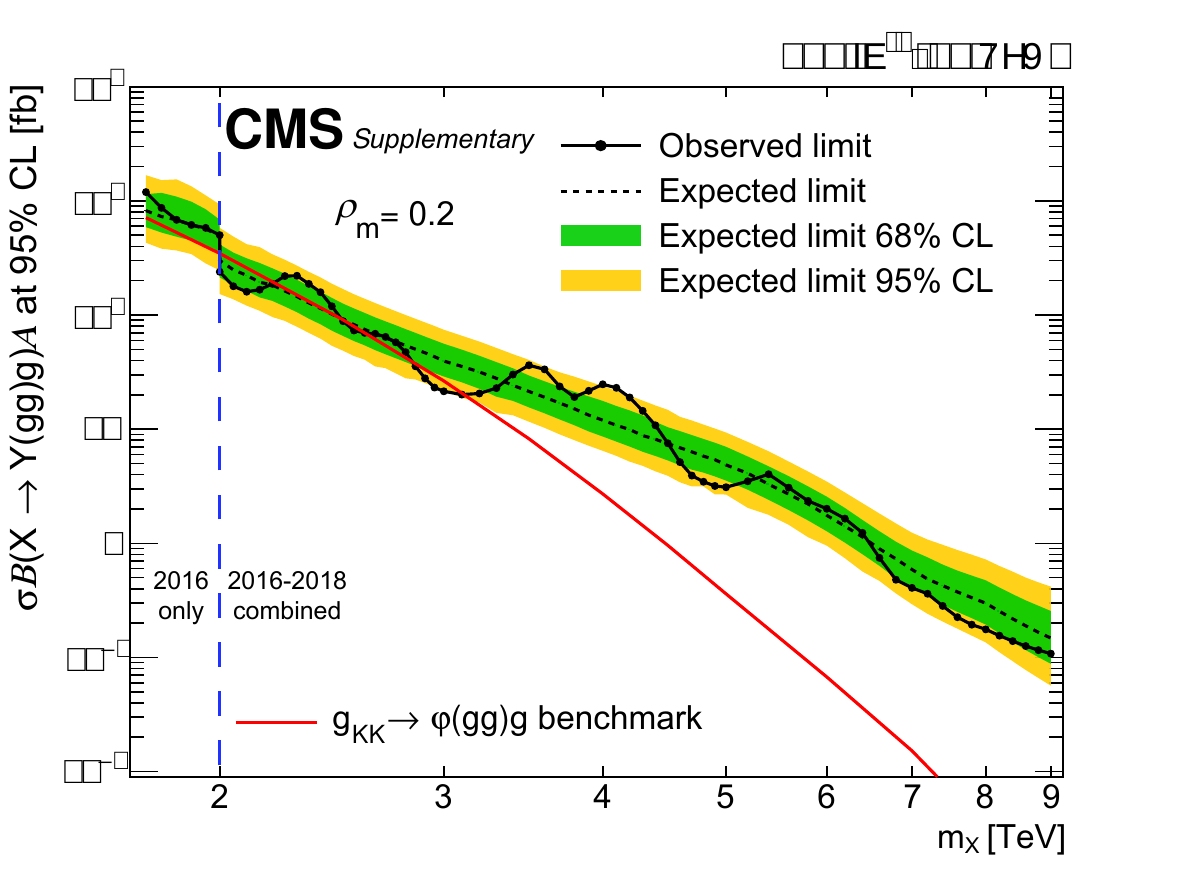}
    \includegraphics[width=0.4\textwidth]{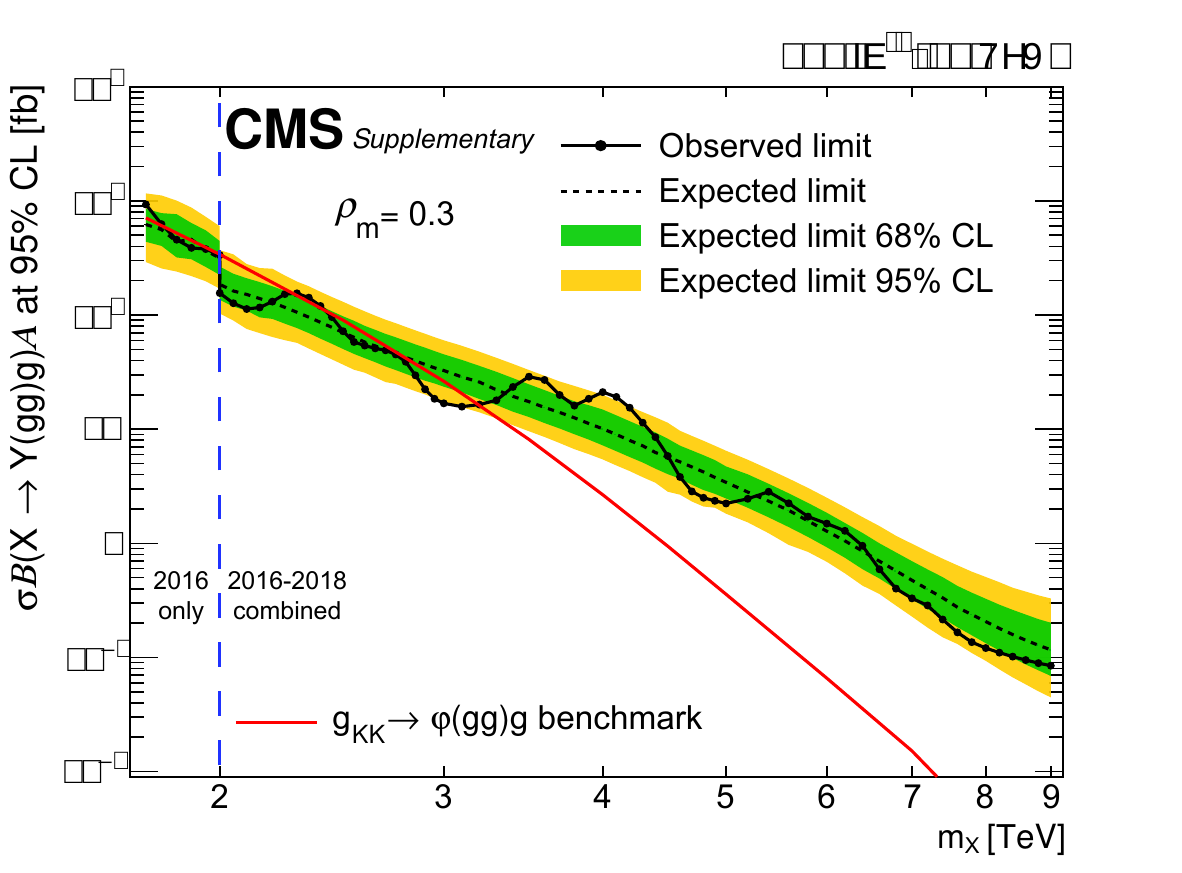}
    \includegraphics[width=0.4\textwidth]{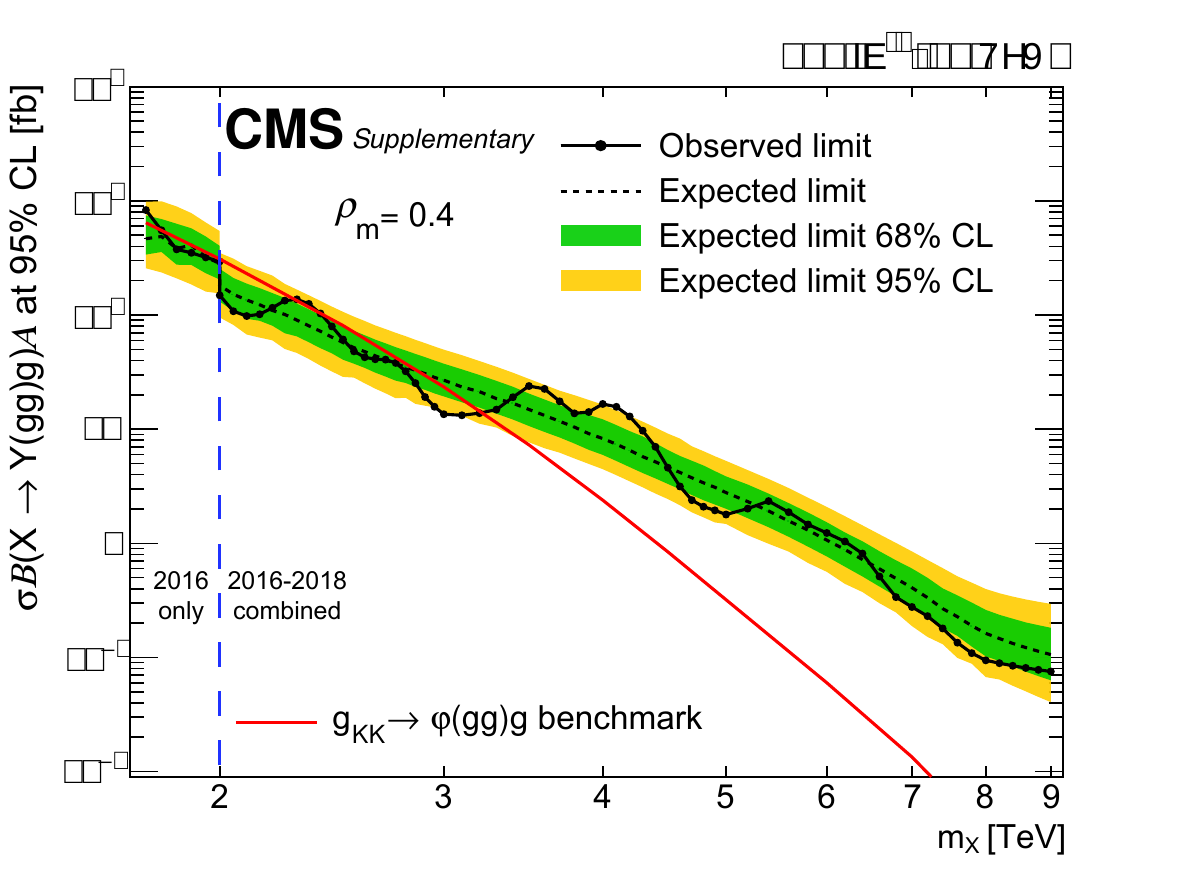}
    \includegraphics[width=0.4\textwidth]{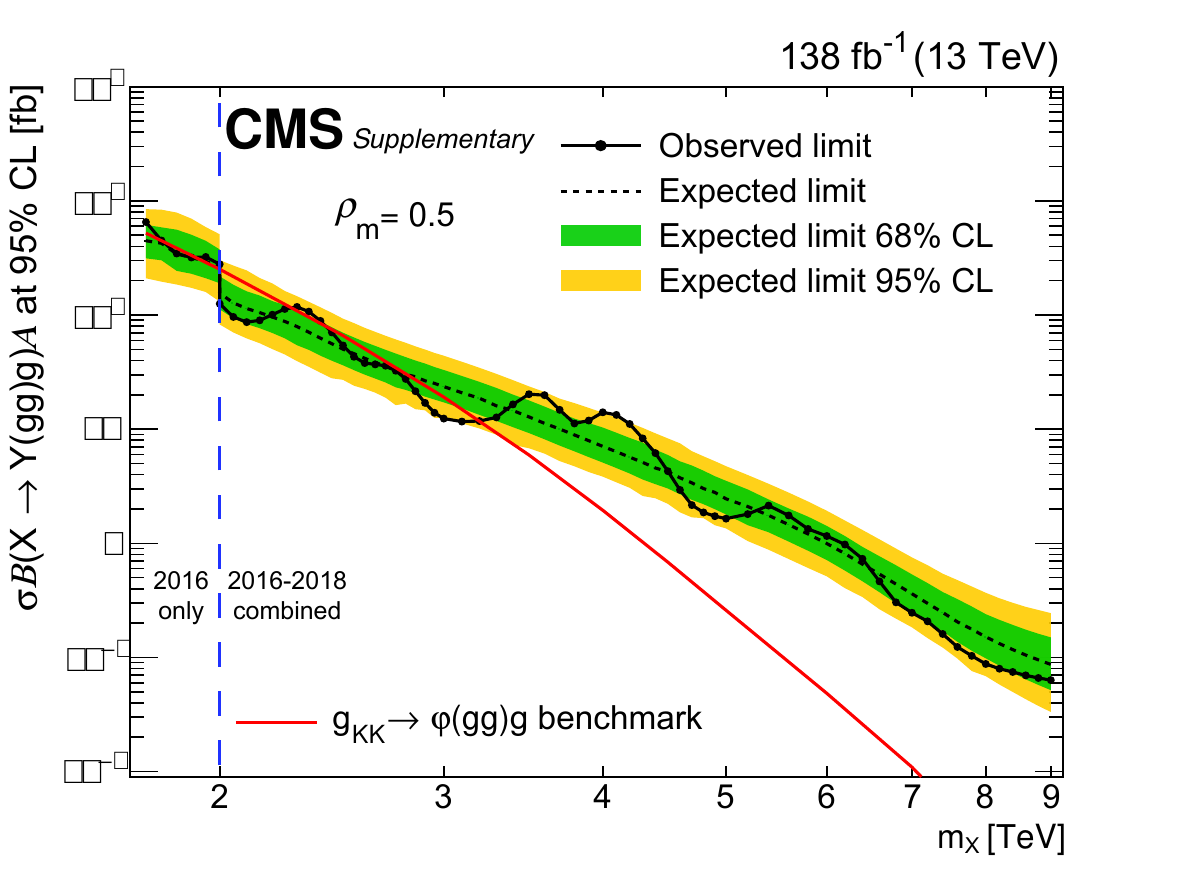}
    \includegraphics[width=0.4\textwidth]{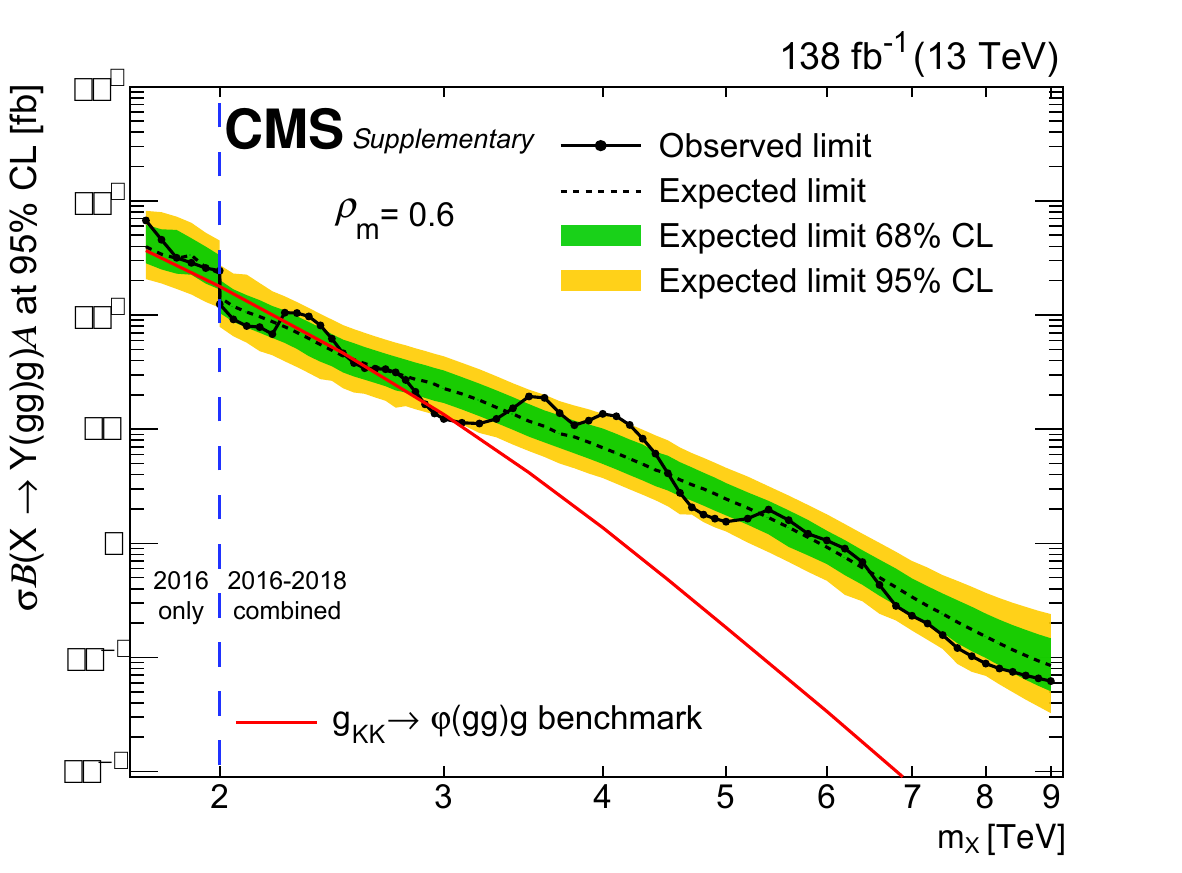}
    \includegraphics[width=0.4\textwidth]{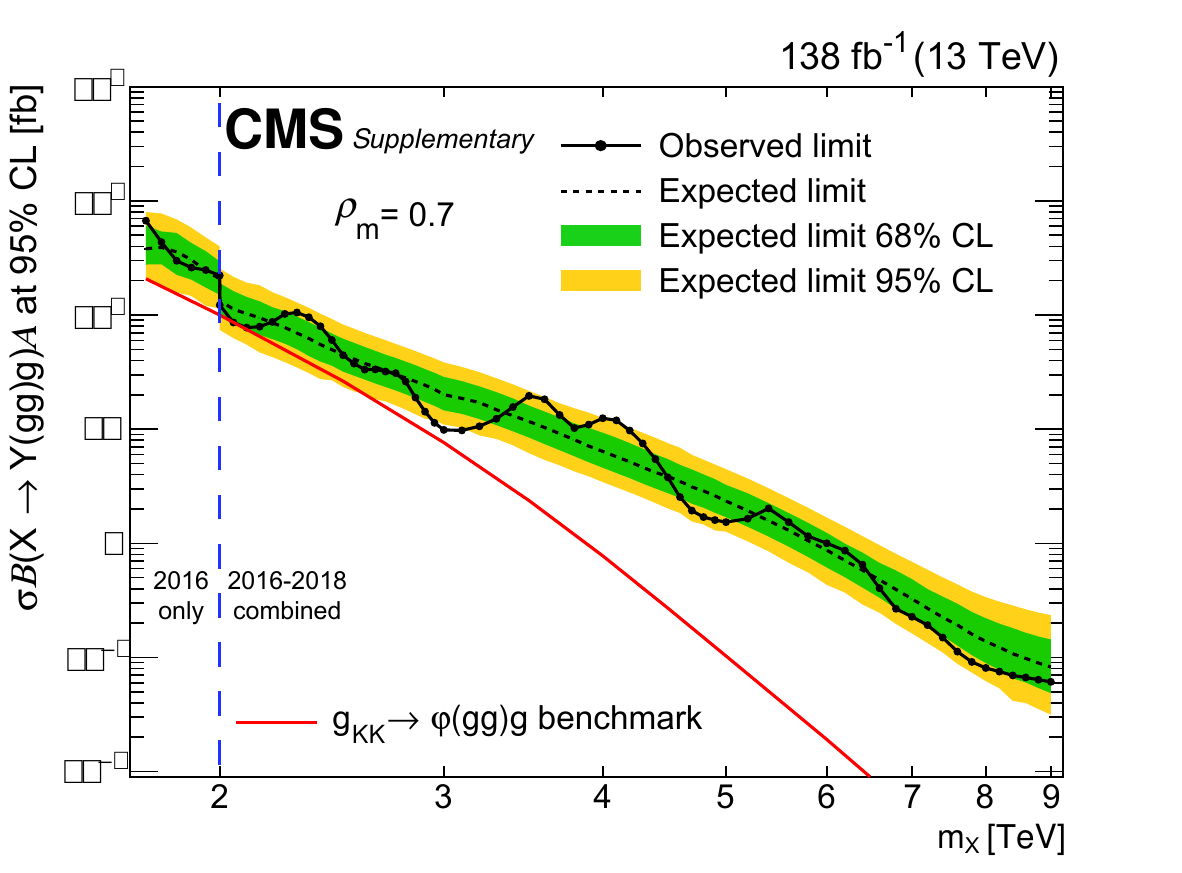}
    \includegraphics[width=0.4\textwidth]{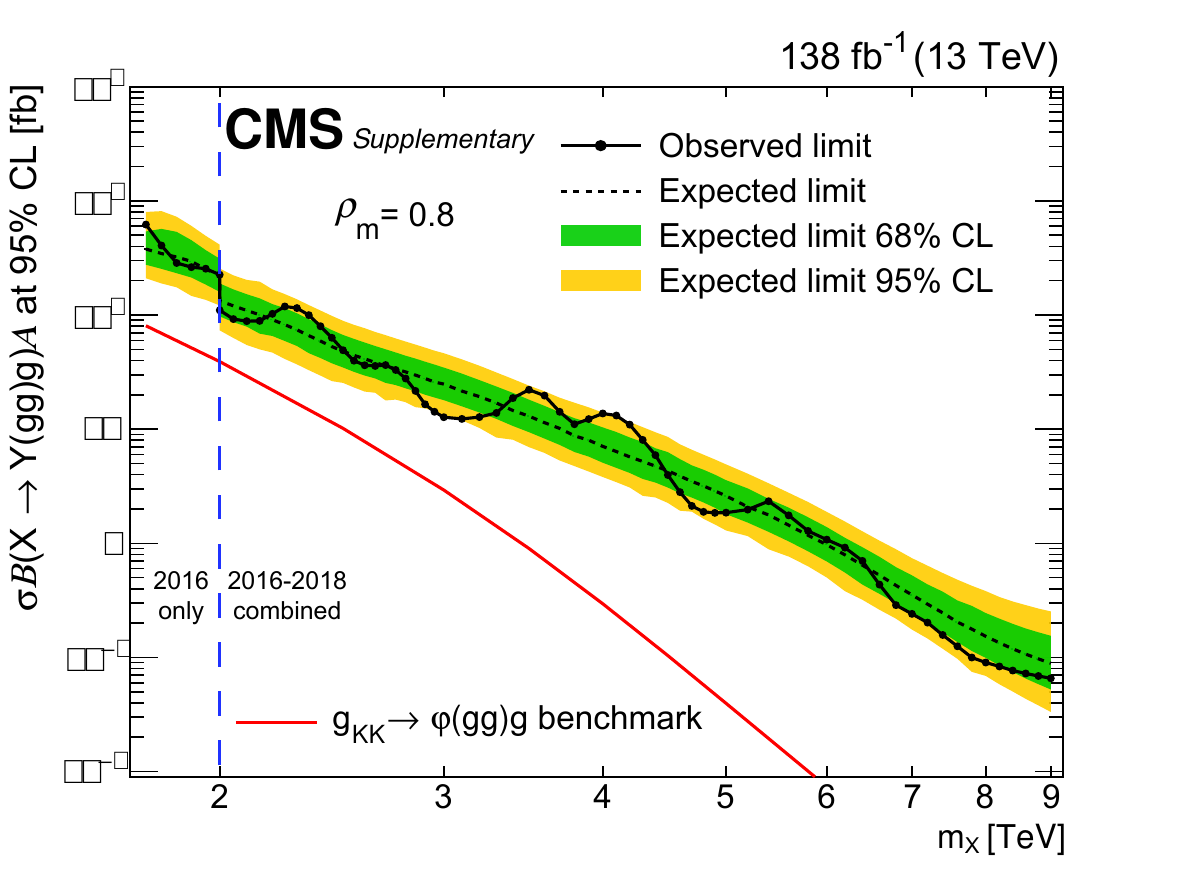}
    \caption{Limits at 95\% \CL on $\sigma\mathcal{B}(\PX\to\PY(\Pg\Pg)\Pg)\mathcal{A}$ for resonances $\PX$ in different $\rho_{m}$ scenarios. Only 2016 data is used to derive limits below 2.0\TeV because of the higher trigger thresholds in 2017 and 2018. Theoretical predictions are also shown for the benchmark $\PGKK$ model.}
    \label{fig:limits_ggg}
\end{figure*}

\begin{figure*}[htpb!]
    \centering
    \includegraphics[width=0.4\textwidth]{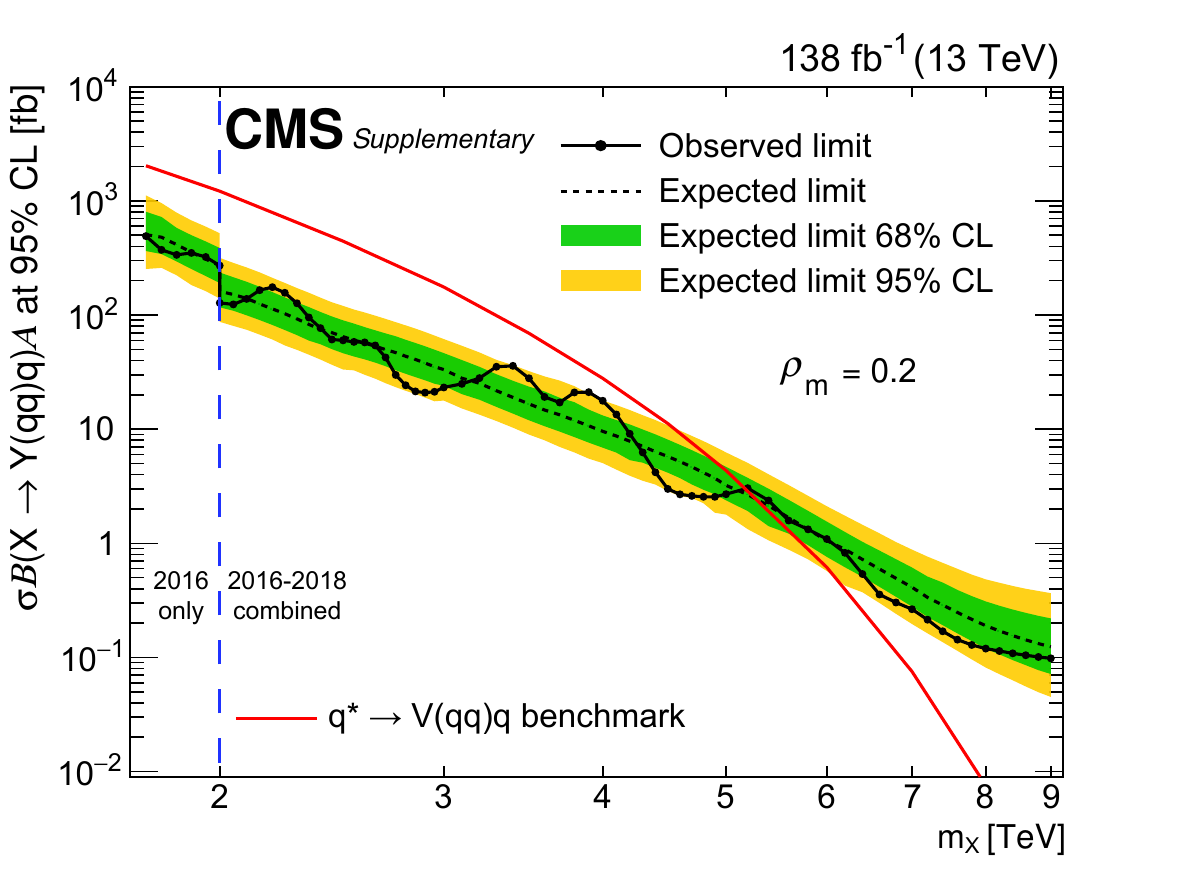}
    \includegraphics[width=0.4\textwidth]{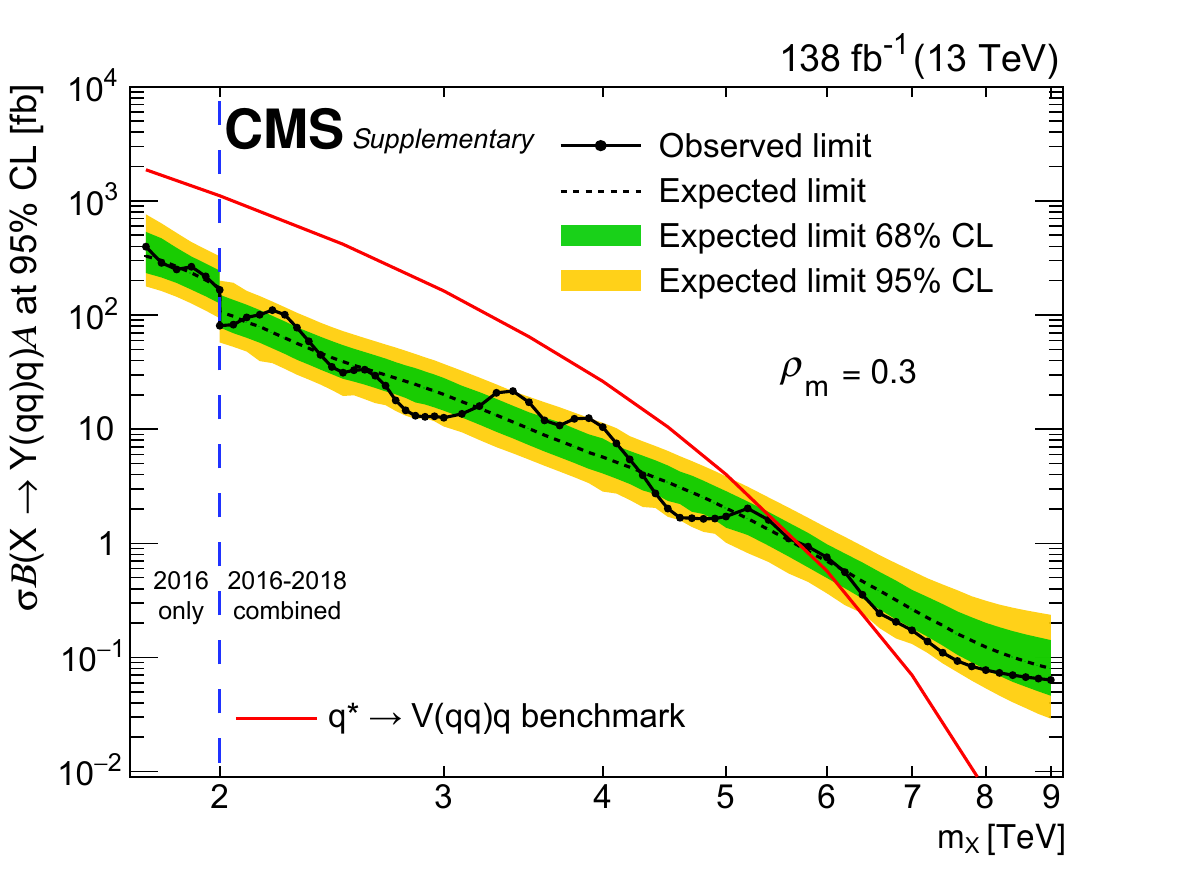}
    \includegraphics[width=0.4\textwidth]{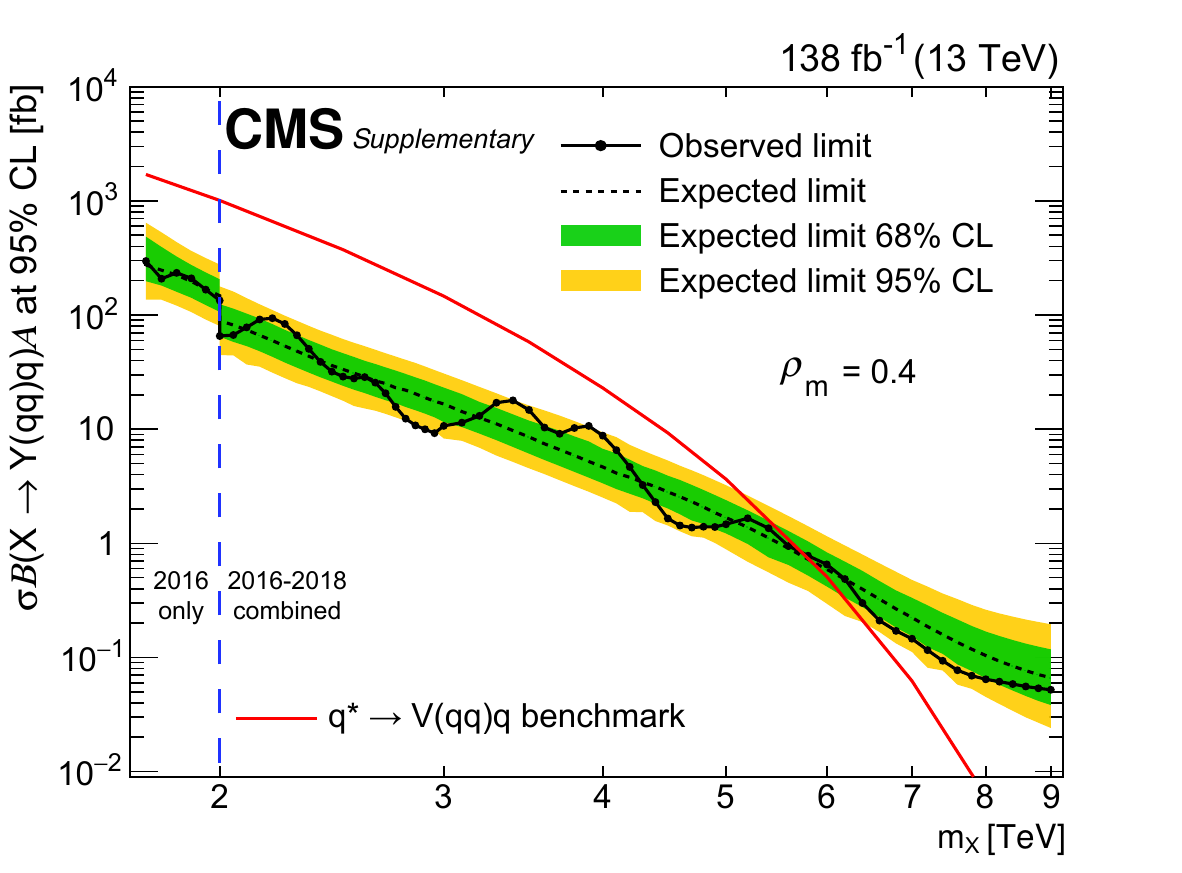}
    \includegraphics[width=0.4\textwidth]{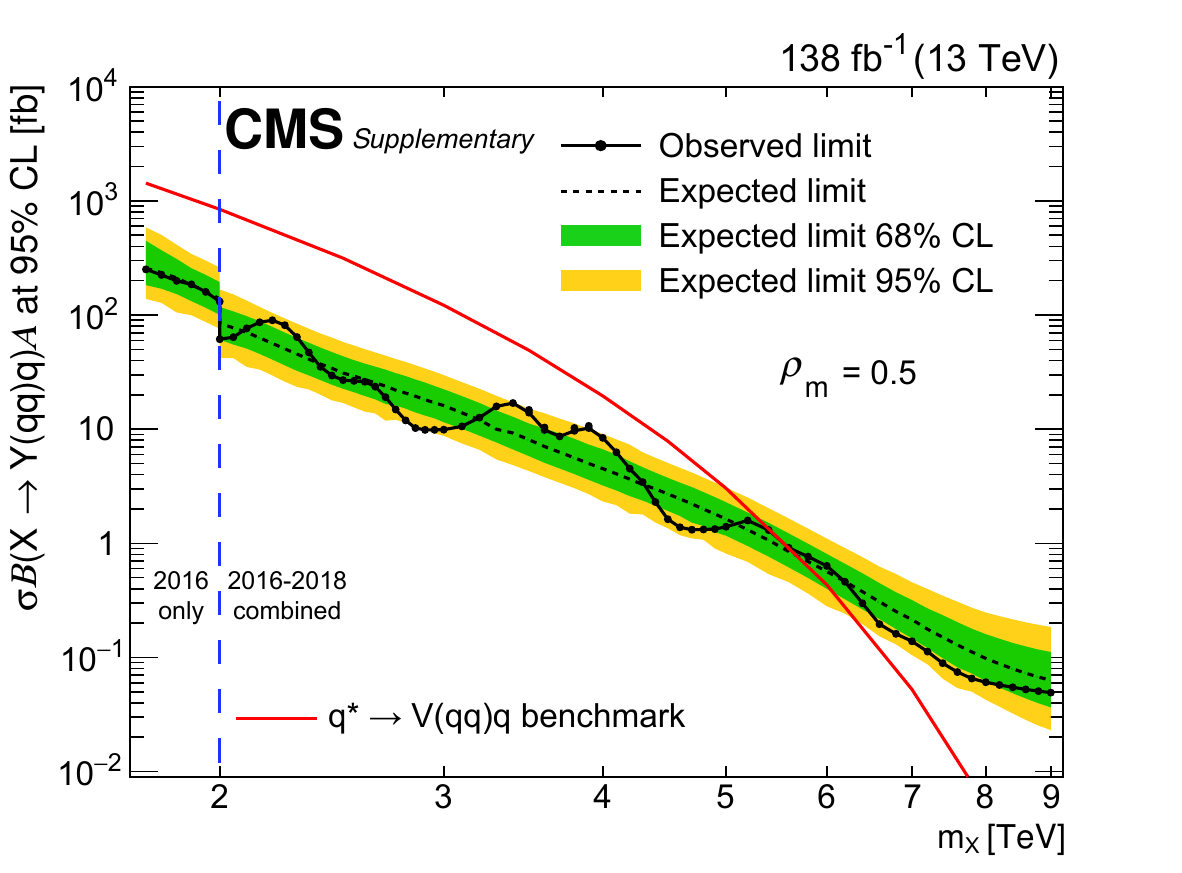}
    \includegraphics[width=0.4\textwidth]{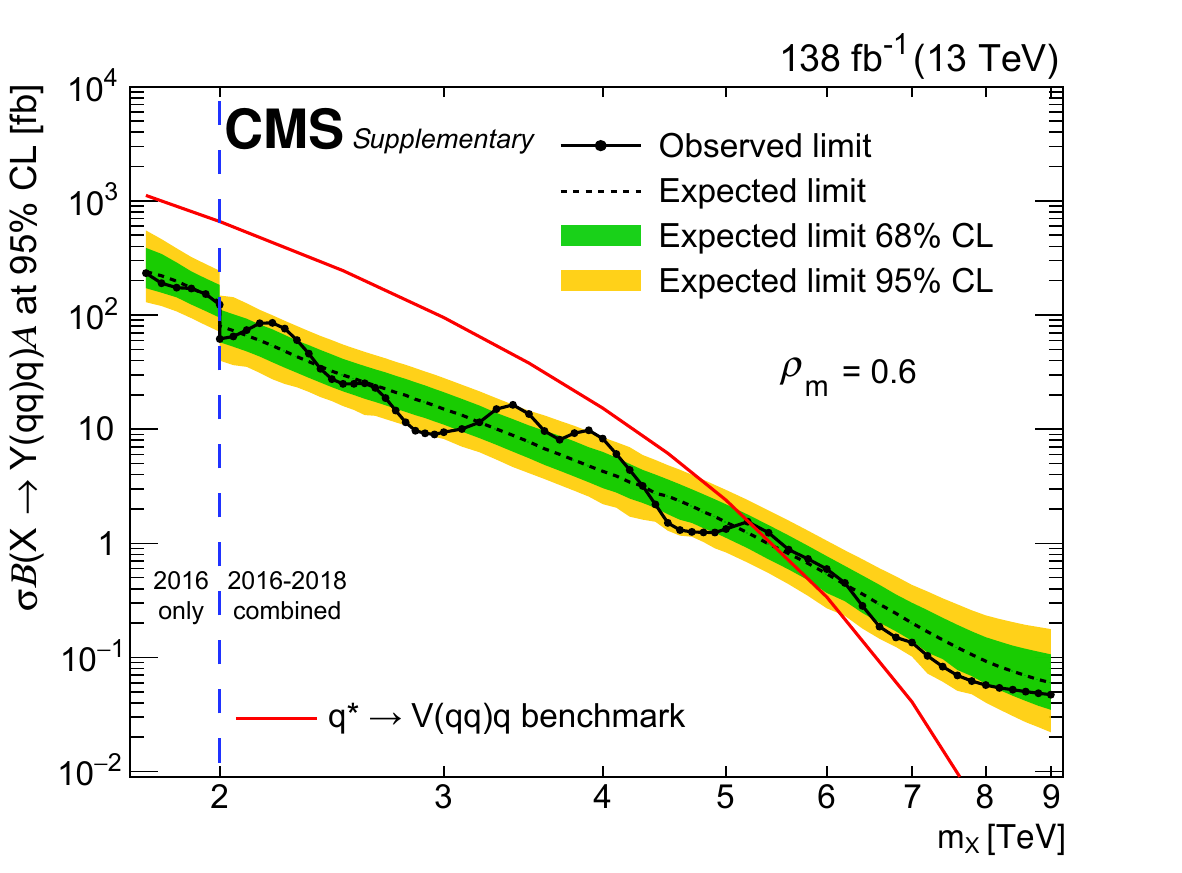}
    \includegraphics[width=0.4\textwidth]{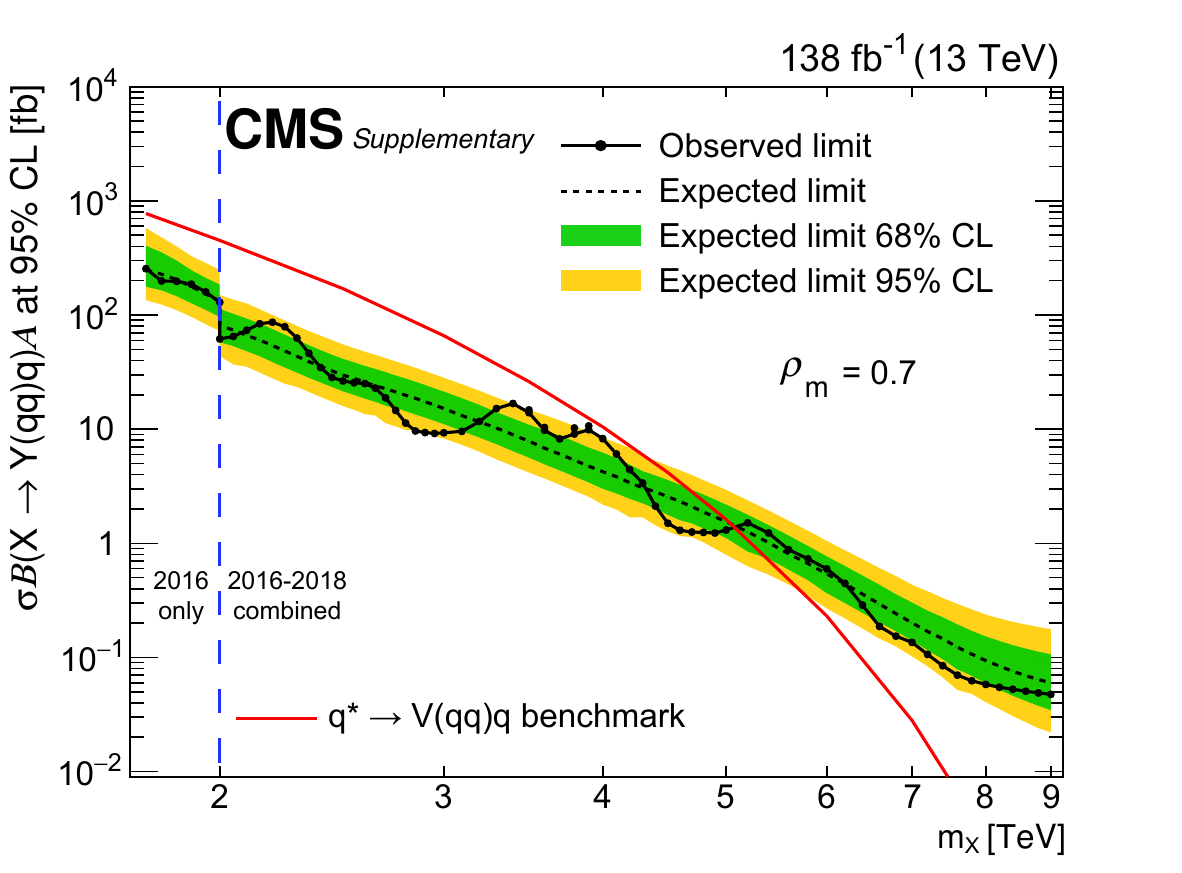}
    \includegraphics[width=0.4\textwidth]{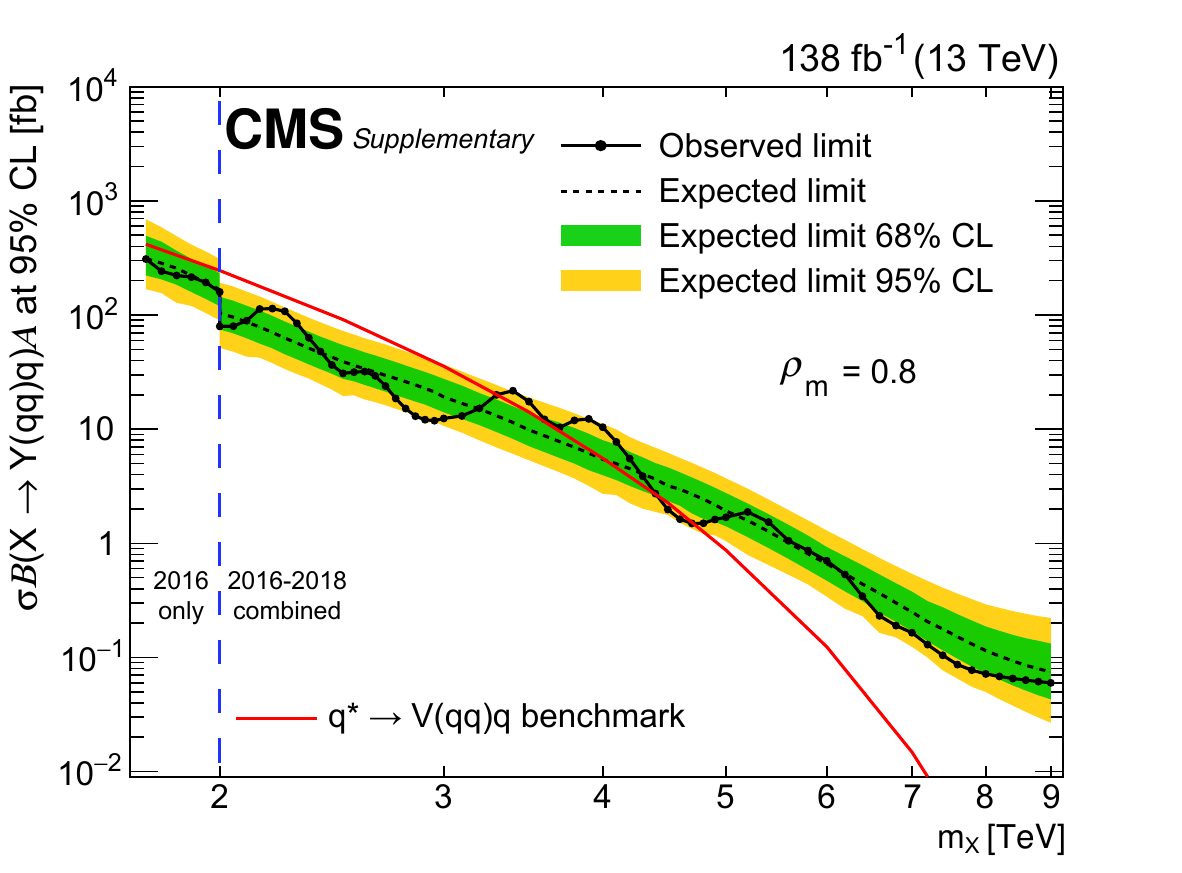}
    \caption{Limits at 95\% \CL on $\sigma\mathcal{B}(\PX\to\PY(\PQq\PQq)\PQq)\mathcal{A}$ for resonances $\PX$ in different $\rho_{m}$ scenarios. Only 2016 data is used to derive limits below 2.0\TeV because of the higher trigger thresholds in 2017 and 2018. Theoretical predictions are also shown for the benchmark $\PQq^{*}$ model.}
    \label{fig:limits_qqq}
\end{figure*}

\begin{figure*}[htpb!]
    \centering
    \includegraphics[width=0.7\textwidth]{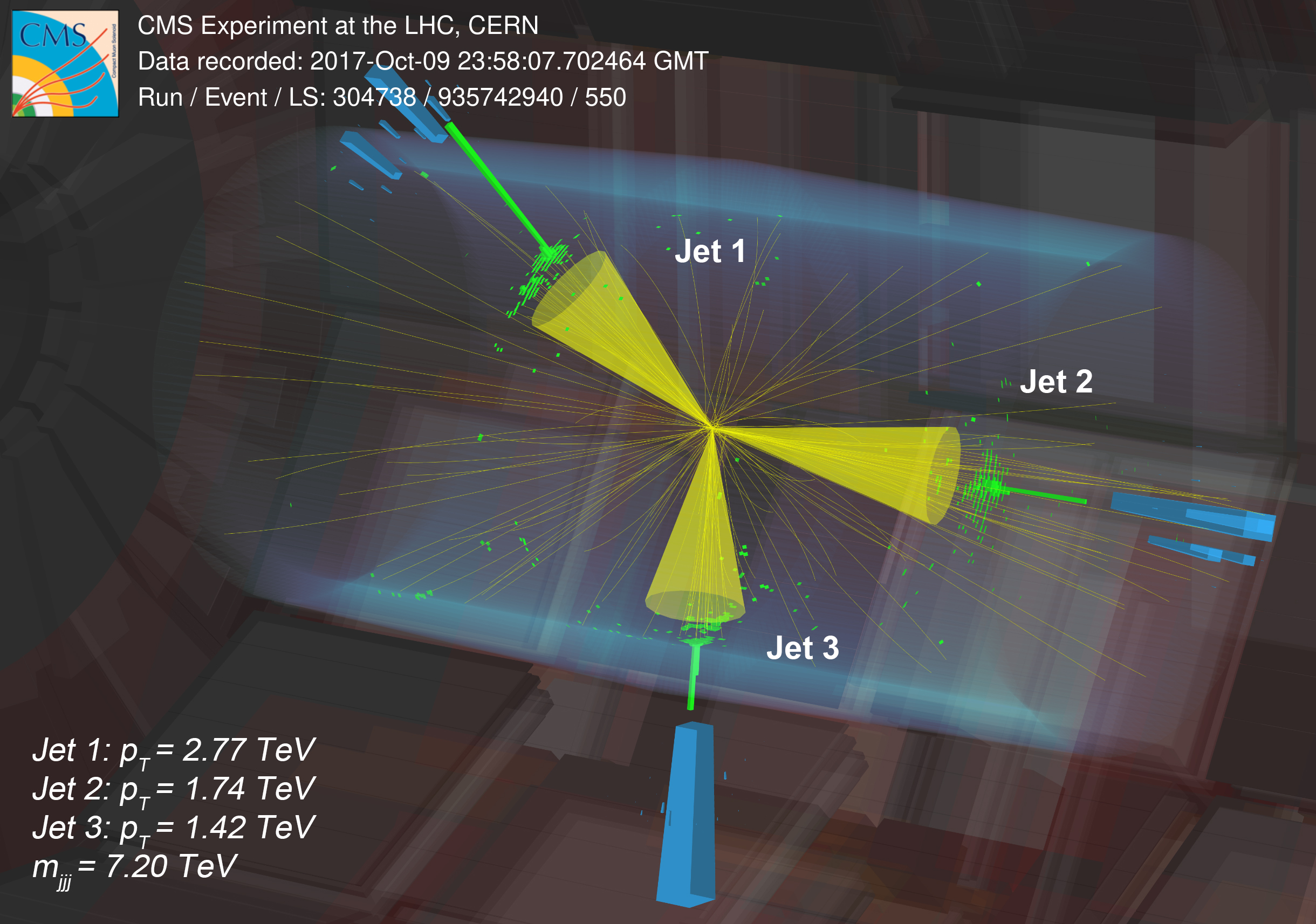}
    \caption{Three-dimensional display of the event with the highest $\mjjj$ of 7.20\TeV. Energy deposited in the electromagnetic (green) and hadronic (blue) calorimeters and the reconstructed tracks of charged particles (yellow) are shown. Reconstructed three most energetic jets are represented by the yellow cones.}
    \label{fig:event_display_1}
\end{figure*}
}\cleardoublepage \section{The CMS Collaboration \label{app:collab}}\begin{sloppypar}\hyphenpenalty=5000\widowpenalty=500\clubpenalty=5000
\cmsinstitute{Yerevan Physics Institute, Yerevan, Armenia}
{\tolerance=6000
A.~Hayrapetyan, A.~Tumasyan\cmsAuthorMark{1}\cmsorcid{0009-0000-0684-6742}
\par}
\cmsinstitute{Institut f\"{u}r Hochenergiephysik, Vienna, Austria}
{\tolerance=6000
W.~Adam\cmsorcid{0000-0001-9099-4341}, J.W.~Andrejkovic, T.~Bergauer\cmsorcid{0000-0002-5786-0293}, S.~Chatterjee\cmsorcid{0000-0003-2660-0349}, K.~Damanakis\cmsorcid{0000-0001-5389-2872}, M.~Dragicevic\cmsorcid{0000-0003-1967-6783}, A.~Escalante~Del~Valle\cmsorcid{0000-0002-9702-6359}, P.S.~Hussain\cmsorcid{0000-0002-4825-5278}, M.~Jeitler\cmsAuthorMark{2}\cmsorcid{0000-0002-5141-9560}, N.~Krammer\cmsorcid{0000-0002-0548-0985}, D.~Liko\cmsorcid{0000-0002-3380-473X}, I.~Mikulec\cmsorcid{0000-0003-0385-2746}, J.~Schieck\cmsAuthorMark{2}\cmsorcid{0000-0002-1058-8093}, R.~Sch\"{o}fbeck\cmsorcid{0000-0002-2332-8784}, D.~Schwarz\cmsorcid{0000-0002-3821-7331}, M.~Sonawane\cmsorcid{0000-0003-0510-7010}, S.~Templ\cmsorcid{0000-0003-3137-5692}, W.~Waltenberger\cmsorcid{0000-0002-6215-7228}, C.-E.~Wulz\cmsAuthorMark{2}\cmsorcid{0000-0001-9226-5812}
% ...
\par}
\vskip\cmsinstskip
\dag:~Deceased\\
$^{1}$Also at Yerevan State University, Yerevan, Armenia\\
$^{2}$Also at TU Wien, Vienna, Austria\\
% ...
\end{sloppypar}
%%% END EDITABLE REGION %%%
% skeleton_end
\end{document}